\pdfoutput=1
\RequirePackage{etex}
\documentclass{aa}  

\usepackage{multirow} 
\usepackage[normalem]{ulem} 
\useunder{\uline}{\ul}{} 

\usepackage{xcolor}
\usepackage{color} 
\usepackage[table,xcdraw]{xcolor} 
\usepackage{lmodern}
\usepackage{mathtools}
\usepackage{sidecap}
\usepackage{lineno}
\usepackage{amsmath, amssymb, amsfonts}
\usepackage{units}
\usepackage{subfig}
\usepackage{natbib}
\usepackage{booktabs}
\usepackage{tabularx}
\usepackage{makecell}

\usepackage{appendix}
\usepackage{graphicx}
\usepackage{txfonts}
\usepackage[hypertexnames=false]{hyperref}
\begin{document} 

   \title{Infrared-enhanced Photometric Redshifts for the Dark Energy Survey Y6 Gold catalogue}

   \author{M.~M.~Puebla\inst{1,2,3}
          \and
          I.~Sevilla-Noarbe\inst{3}
          \and
          J.~de~Vicente\inst{3}
          \and
          L.~Toribio San Cipriano\inst{3}
          \and
          J.~Carretero\inst{3,4}
          \and
          A.~Drlica-Wagner\inst{5,6,7}
          \and
          J.García-Bellido\inst{8}
          \and
          D.~Gruen\inst{9,10}
          \and
          J.~Gschwend\inst{11}
          \and
          T.A.~Manning\inst{12}
          \and
          M.~de~la~Osa\inst{3}
          \and
          A.~Porredon\inst{3}
          \and
          N.~Reynes\inst{3}
          \and
          E.~Sánchez\inst{3}
          \and
          P.~Tallada\inst{3,4}
          \and
          N.~Weaverdyck\inst{13,14}
          }

   \institute{Departamento de Física de la Tierra y Astrofísica, Fac. de C.C. Físicas, Universidad Complutense de Madrid (UCM), E-28040 Madrid, Spain \\
    \email{marinm62@ucm.es}
    \and
    Instituto de Física de Partículas y del Cosmos, IPARCOS, Fac. C.C. Físicas, Universidad Complutense de Madrid, E-28040 Madrid, Spain
    \and
    Centro de Investigaciones Energéticas, Medioambientales y Tecnológicas (CIEMAT), Avda. Complutense 40, E-28040, Madrid, Spain\\
    \email{ignacio.sevilla@ciemat.es}
    \and
    Port d'Informació Científica (PIC), Campus UAB, C. Albareda s/n, 08193 Bellaterra, Barcelona, Spain
    \and
    Department of Astronomy and Astrophysics, University of Chicago, Chicago, IL 60637, USA
    \and
    Fermi National Accelerator Laboratory, P. O. Box 500, Batavia, IL 60510, USA
    \and
    Kavli Institute for Cosmological Physics, University of Chicago, Chicago, IL 60637, USA
    \and
    Instituto de Fisica Teorica UAM/CSIC, Universidad Autonoma de Madrid, 28049 Madrid, Spain
    \and
    University Observatory, Faculty of Physics, Ludwig-Maximilians-Universität, Scheinerstr. 1, 81679 Munich, Germany
    \and
    Excellence Cluster ORIGINS, Boltzmannstr. 2, 85748 Garching, Germany
    \and
    Laborat\'orio Interinstitucional de e-Astronomia - LIneA, Av. Pastor Martin Luther King Jr, 126 Del Castilho, Nova Am\'erica Offices, Torre 3000/sala 817 CEP: 20765-000, Brazil
    \and
    Center for Astrophysical Surveys, National Center for Supercomputing Applications, University of Illinois Urbana-Champaign, 1205 West Clark St., Urbana, IL 61801, USA
    \and
    Lawrence Berkeley National Laboratory, 1 Cyclotron Road, Berkeley, CA 94720, USA
    \and
    Berkeley Center for Cosmological Physics, Department of Physics, University of California, Berkeley, CA 94720, US
}

 
  \abstract
   {The Dark Energy Survey (DES) provides optical data across 5000 square degrees of the southern sky, enabling a broad range of extragalactic and cosmological studies. Combining DES data with infrared surveys offers the opportunity to improve its photometric redshift (photo-z) estimates. }
   {To investigate improvements in photometric redshift estimation achieved by combining DES optical data with infrared measurements from the VISTA Hemisphere Survey (VHS) and the Wide-field Infrared Survey Explorer (WISE), and release an updated version of the catalogue.}
   {We performed a positional sky cross-match between the DES Y6 Gold catalogue matched to a spectroscopic dataset, the 2013 AllWISE Data Release, and VHS Data Release 5, in order to test these improvements using the Directional Neighbourhood Fitting (DNF) algorithm (Y6 Gold catalogue reference estimator). We additionally matched it to the unWISE catalogue to verify the performance against this deeper dataset.}
   {Adding infrared data reduces all the metrics (scatter, bias and outlier fraction) in photo-$z$ estimates, particularly at higher redshifts in comparison with only using optical data from DES. The obtained results are globally better for the DES+WISE sample, with improvements that are statistically significant. On the other hand, the addition of the VHS bands to available depth is only marginal.} 
   {The combined use of DES and WISE W1 and W2 data improves the photometric redshift metrics analysed here. The addition of VHS data at the DES and VHS depths explored here, does not provide any further improvement at $z < 1.5$, indicating that, under these constraints, WISE data may already capture the key infrared features and depth needed for accurate photo-$z$ estimation. In addition, low signal-to-noise ($<10$) infrared data does not contribute to any improvement beyond the DES optical dataset.}


   \maketitle
%

\section{Introduction}

    Photometric redshift estimation is an essential ingredient for many cosmological probes, such as galaxy clustering or weak gravitational lensing analyses. Galaxy surveys such as the Dark Energy Survey (DES; \cite{Sanchez}), the Vera C. Rubin Observatory Legacy Survey of Space and Time (LSST; \cite{LSST}), and \textit{Euclid} \citep{EUCLID} rely critically on accurate and precise photometric redshift (photo-$z$) estimates over hundreds of millions of galaxies. 
   
   DES has delivered deep, homogeneous optical imaging over $\sim$5000 deg$^2$ in five broad bands ($grizY$), enabling the construction of large galaxy samples with well-characterised photometry \citep{DR2DES}. The recent Year 6 Gold catalogue \citep{y6gold} represents the most complete and refined DES dataset to date, incorporating improved photometric calibration, object classification, and uniformity across the footprint. However, the limited wavelength coverage of optical-only surveys introduces degeneracies in galaxy spectral energy distributions (SEDs), particularly at intermediate and high redshifts, where key spectral features such as the 4000 \AA\ break shift beyond the optical regime.

   A well-established strategy to mitigate these limitations is the inclusion of infrared (IR) photometry, which extends the wavelength baseline and provides additional constraints on galaxy SEDs. Previous studies have demonstrated that near-infrared data can improve photometric redshift performance by reducing scatter and catastrophic outliers (e.g. \cite{Banerji2015}). Similar approaches have been adopted in other wide-field surveys, such as KiDS \citep{KIDS}, where the incorporation of infrared data has contributed to improved cosmological constraints. However, these analyses were typically based on earlier DES data releases, smaller samples, or heterogeneous datasets, and a comprehensive assessment using the full depth and uniformity of the DES Year 6 Gold catalogue is still lacking.
   
   In this work, we present a value-added photometric redshift catalogue based on the DES Year 6 Gold dataset, enhanced with infrared information from the Wide-field Infrared Survey Explorer (WISE; \cite{WISE2010}, \cite{unwise}) and the VISTA Hemisphere Survey (VHS; \cite{VHS}). Using the Directional Neighbourhood Fitting (DNF; \cite{DNF}) algorithm, which is also adopted within the DES collaboration and relevant for future surveys such as LSST, we systematically quantify the impact of infrared data on photo-$z$ performance across redshift and magnitude. This enables us to quantify the relative contribution of different infrared datasets as well, distinguishing between DES+WISE, DES+VHS or all combined.

   The catalogue is being made accessible through the CosmoHub\footnote{\url{https://cosmohub.pic.es/catalogs/414}} platform \citep{cosmohub_ref1,cosmohub_ref2}, enabling efficient access, cross-matching, and integration with other large-scale datasets. This resource is intended to support a wide range of astrophysical and cosmological applications, and to provide a reference for future multiwavelength survey combinations.


\section{Data}
\label{sec:datasets}

\subsection{DES Year 6 Gold catalogue}

    The main subject of this work is the dataset from \textbf{DES \textit{(Dark Energy Survey)}}\footnote{\textcolor{blue}{\href{https://www.darkenergysurvey.org/}{DES.}}}, a project that uses the  Dark Energy Camera (DECam; \cite{DECAM}), on the Victor M. Blanco 4m telescope, situated at the Cerro Tololo Inter-American Observatory (CTIO) in Chile. DES is a wide-field optical survey, which started in 2013 and ended in 2019, that aims to understand the nature and evolution of dark energy. It has five photometric broad-band filters: $g,r,i,z,Y$ extending from 400 to 1060 nm, covering an area of approximately 5000 $deg^2$ in the south Galactic cap, with about ten overlapping dithered exposures per filter (90 seconds for $g,r,i,z$ and 45 seconds for $Y$) to fully map the survey footprint. 

    We study the DES Y6 Gold dataset, a refined compilation from the second data release (DR2; \cite{DR2DES}) of the Dark Energy Survey, which has improved measurements, photometric calibration and object classification. Following quality selections, the resulting reference samples have around 448 million galaxies and 120 million stars. The median point spread function (PSF) full width at half maximum (FWHM) is 1.13, 0.99, 0.90, 0.87, 0.93 arcsec for $g,r,i,z$, and $Y$ band, respectively (\cite{y6gold}). DES avoids the Galactic plane to reduce contamination from foreground stars and the effects of interstellar dust extinction.

    In particular, for our results we work with a spectroscopic sample from Y6 Gold which contains 545796 galaxies, obtained by DECam imaging overlapping data from spectroscopic redshift surveys (\cite{Gschwend_spectroscopic_urveys}), so that both magnitudes and spectra have been measured. We use the coadd\textunderscore object\textunderscore id (unique identifier for the coadded objects), RA (Right Ascension, J2000), DEC (Declination, J2000), Z (spectroscopic redshift), and photometric magnitudes in the $g,r,i,z$, and $Y$ bands along with their associated uncertainty columns. In addition, for the analysis, we select galaxies with $i< 26$, as this cut remains suitable for weak lensing analyses, and photometric errors become very large for fainter magnitudes and make the data useless. In practice most of the matches to spectroscopic data (and infrared surveys later on) have a much brighter cut. This selection additionally has the highest quality cut in spectroscopic quality, according to \texttt{FLAG\_DES} \citep{Gschwend_spectroscopic_urveys}.

    The complete Y6 Gold catalogue is used as well to produce the updated version being made available with this publication.

\subsection{VHS}

We combine DES galaxies with near infrared photometric data from the fifth data release of \textbf{VHS (VISTA Hemisphere Survey)}\footnote{\textcolor{blue}{\href{https://www.eso.org/sci/observing/phase3/data_releases/vhs_dr1.html}{VHS DR1.}}}. VHS is a near-infrared photometric survey which uses the VISTA (\textit{Visible and Infrared Survey Telescope for Astronomy}) telescope at ESO's Cerro Paranal Observatory in Chile. It is equipped with the VIRCAM \textit{(VISTA InfraRed CAMera)}, that enables full coverage of the whole southern celestial hemisphere to 30 times deeper level than the Two Micron All Sky Survey (2MASS; \cite{2mass}) or The Deep Near-Infrared Survey (DENIS; \cite{denis}). The fifth data release (DR5\footnote{\textcolor{blue}{\href{https://archive.eso.org/cms/eso-archive-news/new-data-release-dr5-of-the-eso-public-survey-vista-hemisphere-survey.html}{VHS DR5.}}}) provides a multi-band source catalogue containing 11370 catalogue tiles derived from $Y$, $J$, $H$, and $Ks$ bands observations over eight years, between November 2009 and March 2017. The total sky area covered in at least one photometric band amounts to 16,730 $deg^2$. The point spread function (PSF) of the telescope and camera system, accounting for pixel size, is expected to have a full width at half maximum (FWHM) of approximately 1 arcseconds, with values of 1.03, 1.01, 1.01, and 0.93 arcseconds for the $Y$, $J$, $H$, and $Ks$ bands, respectively.

In this work, we analyse the DES+VHS galaxies obtained by matching the VHS photometric data to the DES spectroscopic sample. We selected the RA, DEC, and Petrosian magnitudes for extended sources in the $J$, and $Ks$ bands along with their associated uncertainty columns, given that $Y$ largely overlaps with the same bands in DES. We found that the addition of the $H$ band was not worthwhile as it imposed a severe limitation in the number of matches if we required it for our analyses, as there was less data acquired for it. 

\subsection{WISE}

For mid-infrared coverage, we use the photometric data from \textbf{WISE\textit{ (Wide-field Infrared Survey Explorer at IPAC)}}\footnote{\textcolor{blue}{\href{https://irsa.ipac.caltech.edu/data/WISE/docs/release/AllWISE/}{AllWISE.}}} (WISE; \cite{WISE2010}). WISE is a wide-field survey of the entire sky in the mid-infrared range, that uses a 40 cm cryogenically cooled space telescope equipped with 62 detectors, and began surveying the sky on 14 January 2010. It has four mid-infrared photometric bands, $W1,W2,W3,W4$, centered at ($3.4, 4.6, 12,22$ $\mathrm{\mu m}$), with angular resolutions of 6.1", 6.4", 6.5", 12.0", respectively. It provides significantly higher sensitivity and coverage than other previous missions such as the InfraRed Astronomical Satellite (IRAS; \cite{IRAS}; \cite{Beichman1988}) or The Infrared Astronomical Mission (AKARI; \cite{Murakami}), enabling the detection of fainter and more distant sources. 

In this work, we use WISE data in two ways, in combination with the matched spectroscopic sample of DES:
    \begin{itemize}
        \item using the official AllWISE catalogue (\cite{2013wise.rept....1C}), which corresponds to observations obtained during the full cryogenic and post-cryogenic \citep{NEOWISE} phases of the mission. From the full catalogue, we selected the RA, DEC, and photometric magnitudes in the $W1$ and $W2$ bands along with their associated uncertainty columns. 
        \item using the unWISE \citep{unwise} unofficial update that incorporates additional data, improves the quality of the coadds and therefore enables the performance of forced photometry on optical detections, for which we use the DECaLS DR10 release\footnote{\url{https://www.legacysurvey.org/dr10/}} \citep{decals}. This data release covers the DES footprint, uses DECam data and produces matched information to forced photometry on these updated WISE images. We selected the RA, DEC, and photometric magnitudes in the $W1,W2$ bands along with their associated uncertainty columns and make the appropriate extinction correction available in this associated catalogue as well. This deeper dataset is used to mass produce the new version of the DES Y6 Gold catalogue, as described below, after comparison against the AllWISE catalogue results.
    \end{itemize}
    
\subsection{Filter response curves}
\begin{figure}[htbp]
 \centering
  \subfloat[DECam filter response curves.]{
   \label{f:filters1}
    \includegraphics[width=0.4\textwidth]{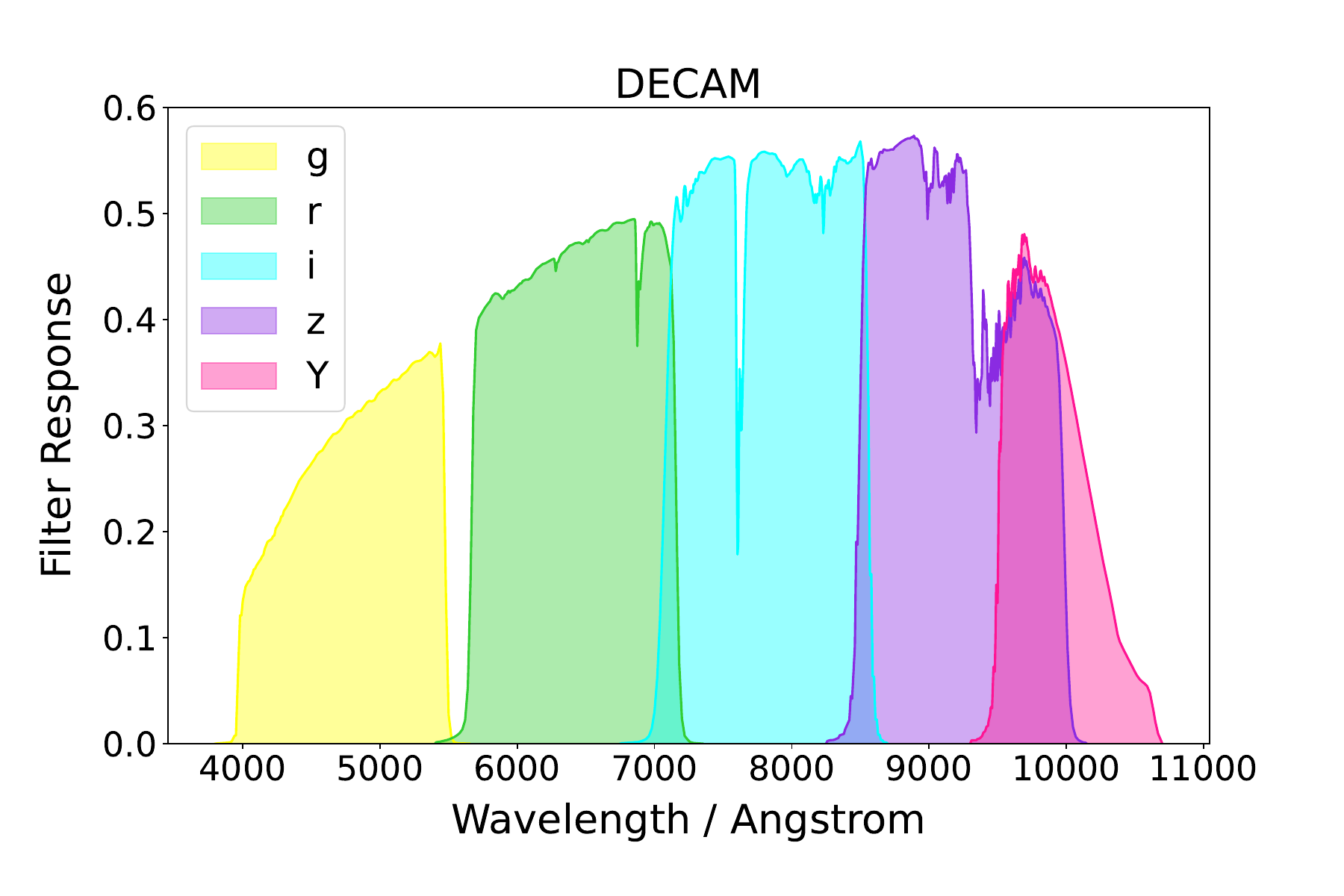}}

  \subfloat[VISTA filter response curves.]{
   \label{f:filters3}    
   \includegraphics[width=0.40\textwidth]{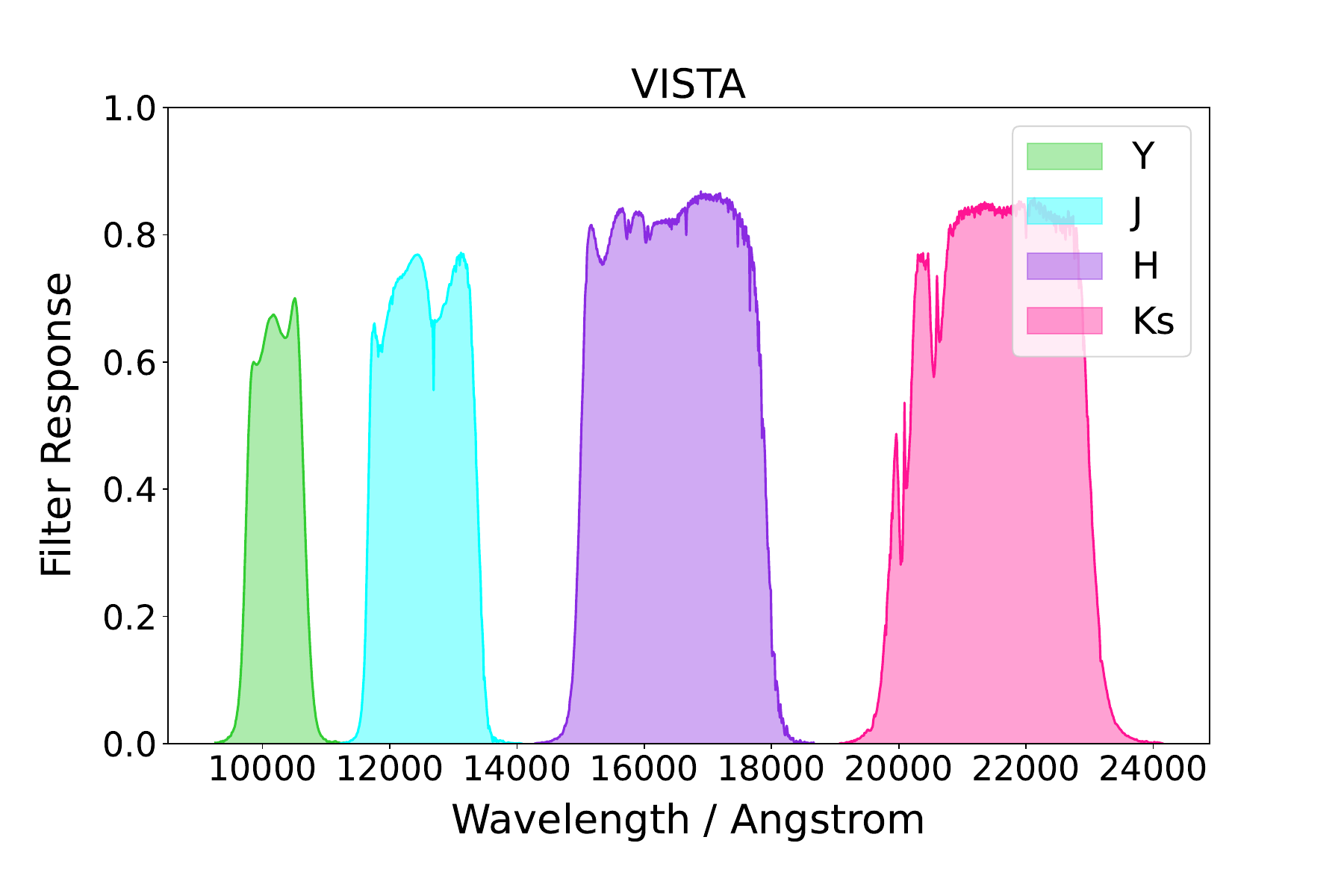}}
%

  \subfloat[WISE filter response curves.]{
   \label{f:filters2}
    \includegraphics[width=0.40\textwidth]{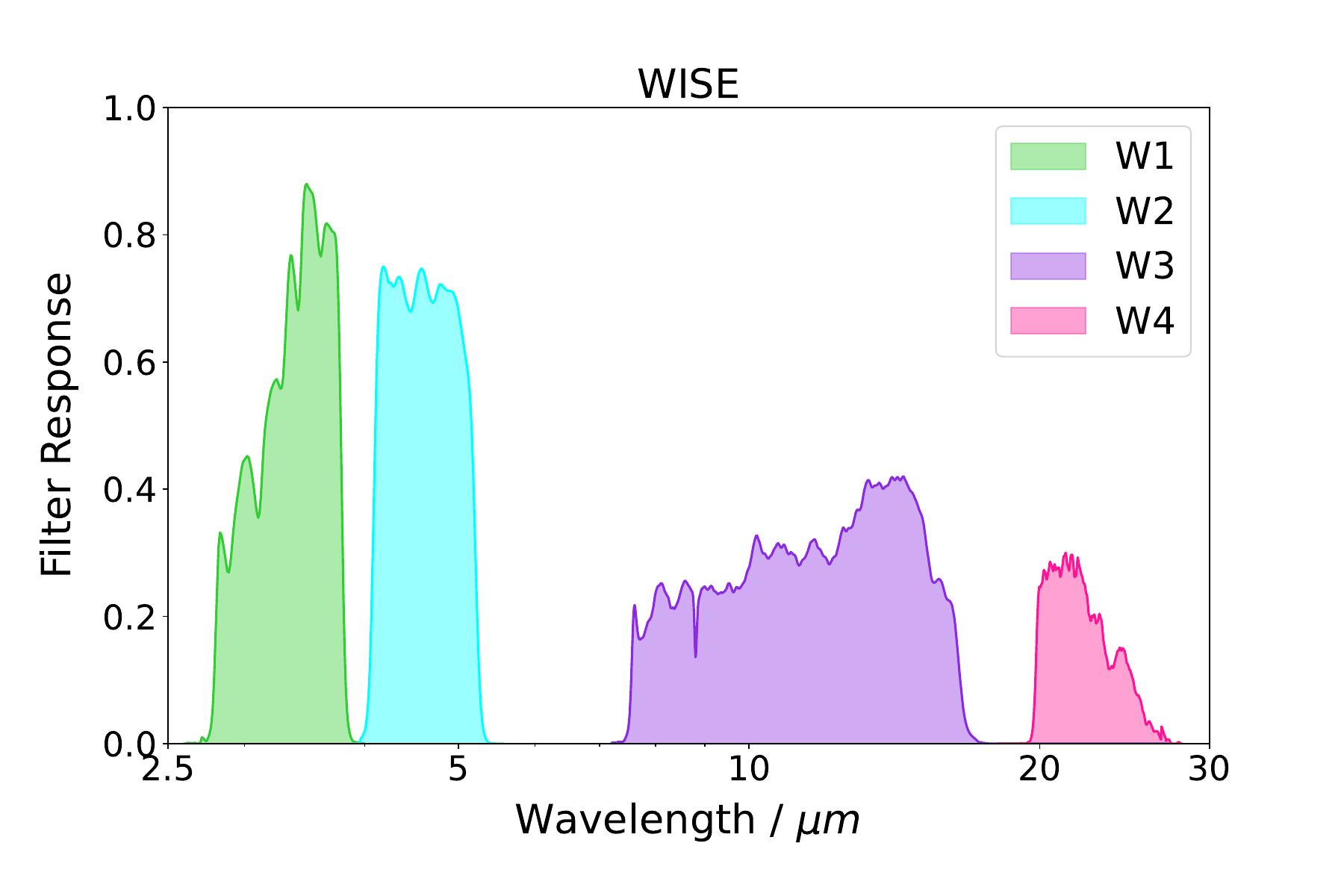}}
 \caption{\centering DECam, VHS and WISE filters.}
  \label{f:filtroscurvastrans}
\end{figure}

    Figure \ref{f:filtroscurvastrans} shows the transmission curves for the $g$, $r$, $i$, $z$, $Y$ filters from DES (\ref{f:filters1}), the $Y$, $J$, $H$, $Ks$ filters from VHS (\ref{f:filters3}) and the $W1$, $W2$, $W3$, $W4$ filters from WISE (\ref{f:filters2}). They illustrate the transmission efficiency of each photometric band used within this work. They are essential for the interpretation of the contribution of different wavelengths to the observed fluxes,  which directly impact on photometric redshift estimation. They take into account instrumental sensitivity, atmospheric effects, and the overall throughput of the system.

\subsection{Cross-match}
    We use the matching and visualization tools from TOPCAT \textit{(Tool for OPerations on Catalogues And Tables)}\footnote{\textcolor{blue}{\href{https://www.star.bris.ac.uk/~mbt/topcat/}{TOPCAT Documentation.}}}(\cite{Topcat}), an  interactive graphical viewer and editor for tabular data. Specifically, the \textit{CDS Upload X-Match} feature was used to create the combined catalogues from the original DES dataset with AllWISE and VHS. The \textit{CDS Upload X-Match} allows matching a local table with external catalogues from the VizieR (\cite{Vizier}) or SIMBAD (\cite{Simbad}) databases, through the CDS X-Match service provided by the Centre de Données astronomiques de Strasbourg (CDS)\footnote{\textcolor{blue}{\href{https://cds.u-strasbg.fr/}{CDS.}}}  

    We performed a positional sky cross-match between the DES spectroscopy training data file, the 2013 AllWISE Data Release, and the VHS Data Release 5. This resulted in the production of three main combined new catalogues: DES+VHS, DES+AllWISE and DES+VHS+AllWISE. These new datasets contain photometric data in 7, 7 and 9 bands, respectively: $grizYW1W2$ (DES+AllWISE), $grizYJKs$ (DES+VHS), and $grizYJKsW1W2$ (DES+VHS+WISE). We match the galaxies in the catalogue of the spectroscopic survey from DES to the galaxies observed photometrically by WISE and VHS, using the positions of the galaxies in the sky plane, with a matching radius of 1". 
 
The choice of this angular tolerance, 1", is supported by an analysis of the median Point Spread Function (PSF) Full Width at Half Maximum (FWHM) of the different photometric bands involved, as presented in Table \ref{tab:seeing}. For DES and VHS, the PSF FWHM ranges from about 0.87$"$ to 1.13$"$, indicating that the sources are quite well-resolved. For WISE, although the spatial resolution is lower (W1: 6.1$"$, W2: 6.4$"$), its photometric sources are typically bright and well-centred, making the match still reliable even at a smaller matching radius like 1$"$.

To ensure that 1$"$ was a reasonable choice for the matching radius, we also tested 1.3$"$ and 1.5$"$ using TOPCAT, and compared the number of matches. The results are shown in Table \ref{tab:BESTALL}. When using the BEST match (the closest source), WISE has a slightly higher number of matched galaxies compared to VHS, but the number of matched galaxies changed very little when increasing the radius for both surveys. This indicates that, for most galaxies, the nearest match is already within 1$"$, and that using a larger radius does not alter the results. For the ALL match (which includes all possible sources within the radius), VHS shows a larger increase in matches compared to its BEST matches. This likely reflects the higher source density and astrometric precision of VHS. 

Based on these results, using a 1$"$ radius is a good choice, as it captures almost all true matches, avoids false ones, and is consistent with the PSF values of the surveys.

\begin{table}[htbp]
\caption{Median PSF FWHM (") from \cite{y6gold}, Pons et al.(2020)\protect\footnotemark\ and \cite{WISE2010}.}
\label{tab:seeing}
\centering
\resizebox{\columnwidth}{!}{
\begin{tabular}{cccccccccc}
\hline\hline
$g$ & $r$ & $i$ & $z$ & $Y_\mathrm{DES}$ & $J$ & $H$ & $K_\mathrm{s}$ & $W1$ & $W2$ \\
\hline
1.13 & 0.99 & 0.90 & 0.87 & 0.93 & 1.01 & 1.01 & 0.93 & 6.1 & 6.4 \\
\hline
\end{tabular}
}
\end{table}

\begin{table}[htbp]
\caption{Number of galaxies with the match angle.}
\label{tab:BESTALL}
\centering
\resizebox{\columnwidth}{!}{
\begin{tabular}{lccc|ccc}
\hline\hline
 & \multicolumn{3}{c|}{BEST} & \multicolumn{3}{c}{ALL} \\
Survey
 & $1\arcsec$ & $1.3\arcsec$ & $1.5\arcsec$
 & $1\arcsec$ & $1.3\arcsec$ & $1.5\arcsec$ \\
\hline
VHS  & 318\,157 & 318\,620 & 319\,027 & 404\,291 & 405\,637 & 407\,796 \\
AllWISE & 377\,225 & 394\,883 & 402\,997 & 377\,225 & 394\,890 & 403\,010 \\
\hline
\end{tabular}
}
\end{table}

\begin{figure}[htbp]
 \centering
  \subfloat[DES.]{
   \label{f:des}   \includegraphics[width=0.3\textwidth]{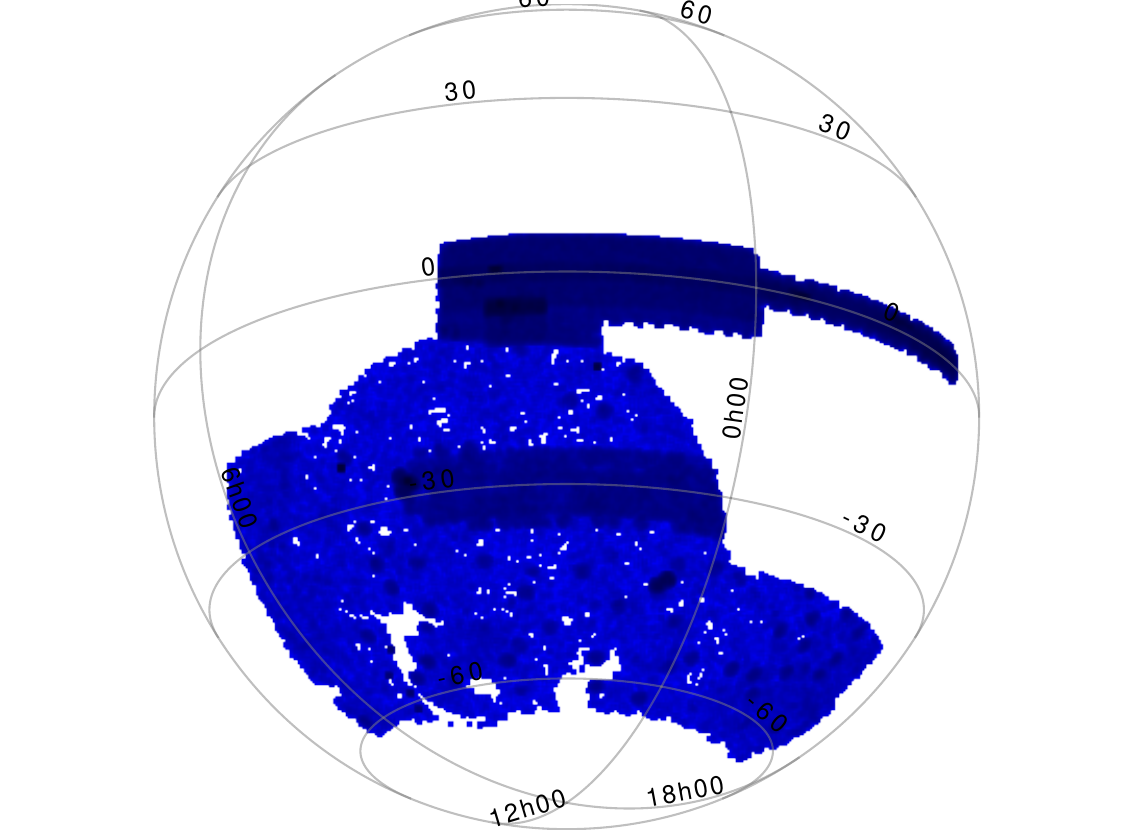}}    

  \subfloat[VHS.]{
   \label{f:vhs} \includegraphics[width=0.3\textwidth]{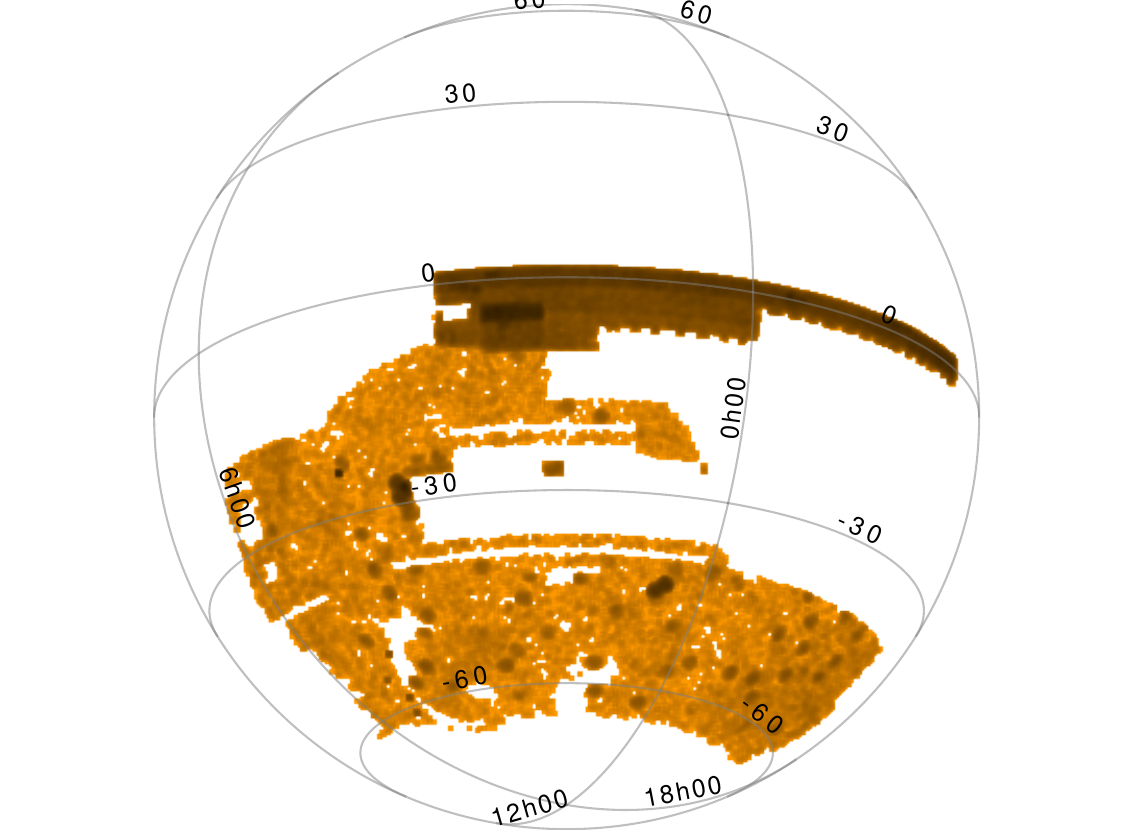}}
    
  \subfloat[AllWISE.]{
   \label{f:WISE}    \includegraphics[width=0.3\textwidth]{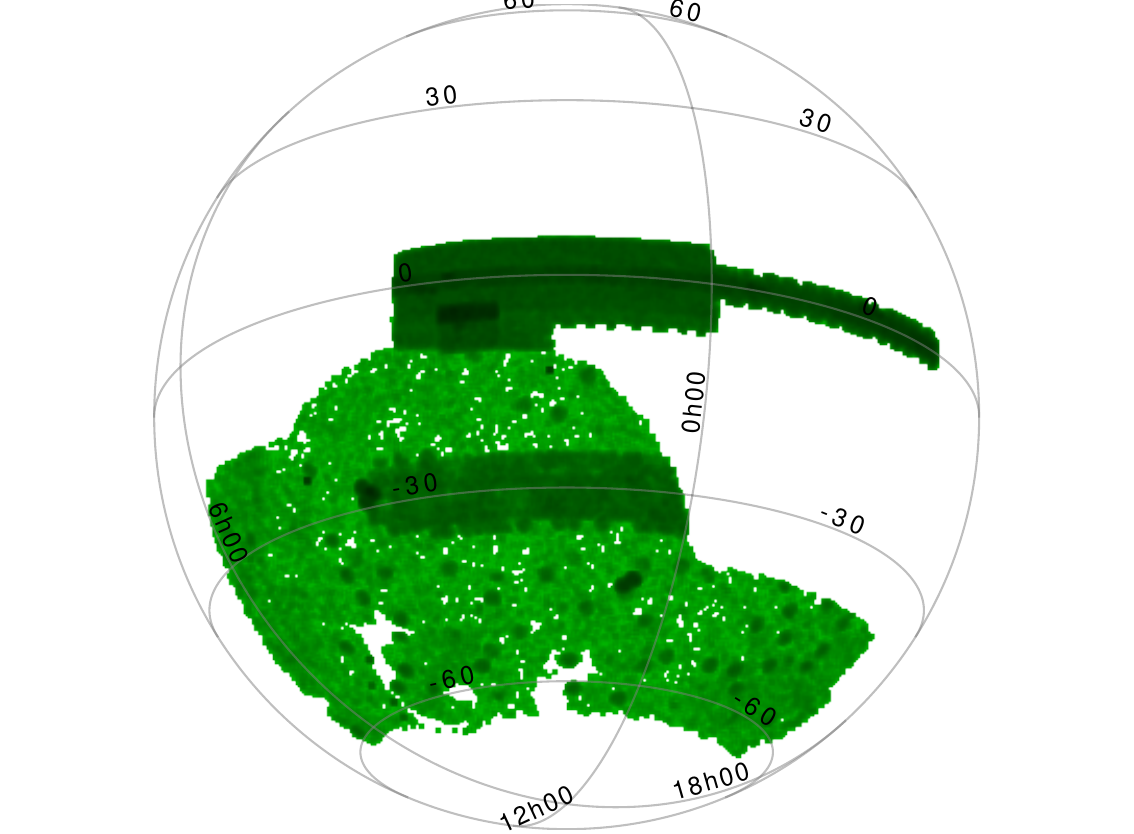}}

  \subfloat[unWISE.]{
   \label{f:unWISE}   \includegraphics[width=0.3\textwidth]{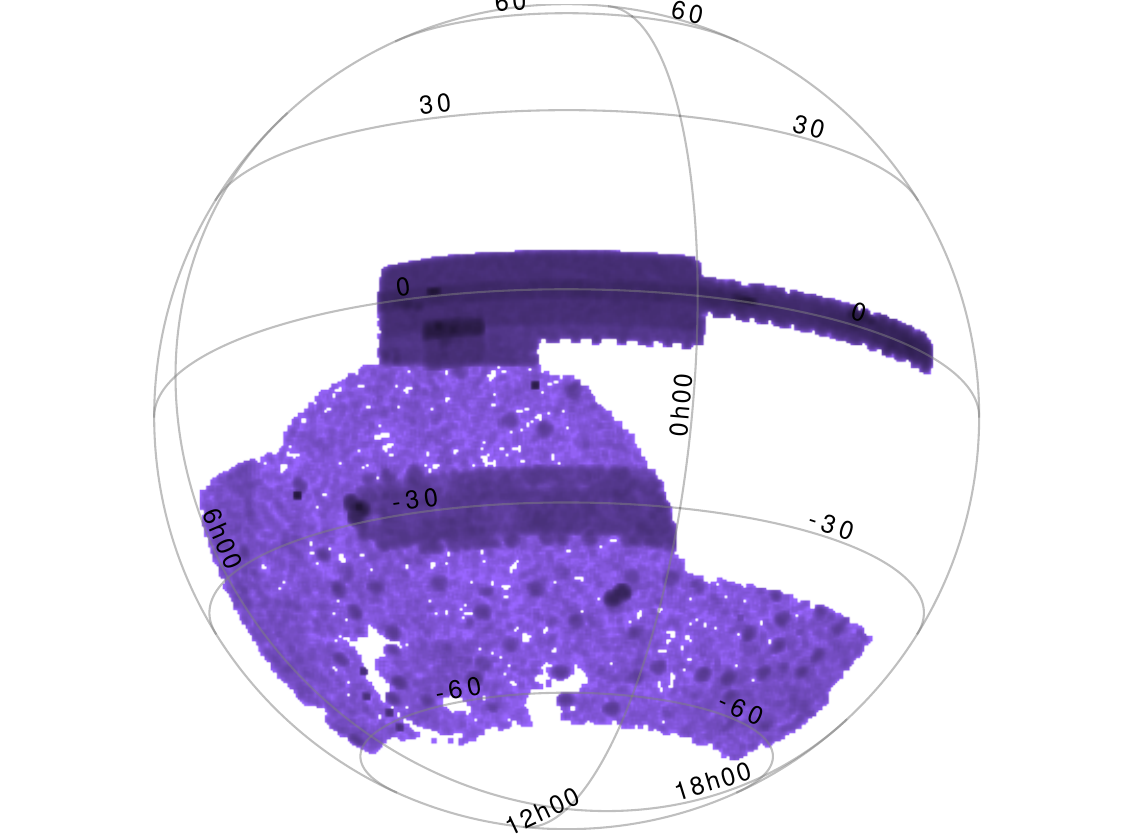}}
 \caption{\centering Distribution of the sources in the matched spectroscopic datasets.}
 \label{f:skymatchcatalogs}
\end{figure}

Figure \ref{f:skymatchcatalogs} shows the sky coverage of the DES spectroscopic dataset and the VHS, AllWISE and unWISE matches (Figures \ref{f:des}, \ref{f:vhs}, \ref{f:WISE}, \ref{f:unWISE} respectively, see also Section \ref{sec:match_complete_gold} for the unWISE case). These plots allow us to visually identify the overlapping regions among the surveys, so that the coverage between the surveys is emphasized.

The number of galaxies in each catalogue is presented in Table \ref{t:numgal}. Some galaxies lacked magnitude information and were therefore excluded from the statistical analysis (reflected in the third column). The reduction in the number of galaxies after cross-matching reflects both the overlapping sources between surveys and the impact of filtering criteria.

\begin{table}[htbp]
\caption{Number of galaxies in the DES spectroscopic sample and in the new merged catalogues before and after applying quality cuts.}
\label{t:numgal}
\centering
\small
\begin{tabular}{lcc}
\hline\hline
Catalogue & Matched galaxies & After basic mag cut \\
\hline
DES(Y6Gold-DR2)        & 545\,796 & 545\,068 \\
DES+VHS (DR5)   & 318\,157 & 281\,349 \\
DES+AllWISE   & 377\,225 & 377\,225 \\
DES+unWISE      & 544\,359 & 511\,504 \\
DES+all         & 254\,243 & \textbf{236\,512} \\
\hline
\end{tabular}
\tablefoot{
The bold-faced entry indicates the subsample used for the comparative analysis. All matches performed at catalog level except the DES+unWISE that uses the DECaLS DR10 forced photometry values over unWISE. DES+all includes a match between the DES spectroscopic catalog, the AllWISE catalog, the VHS DR5 catalog and the unWISE forced photometry association through DECaLS DR10
}
\end{table}

\begin{figure}[htbp]
    \centering
    \includegraphics[width=0.9\linewidth]{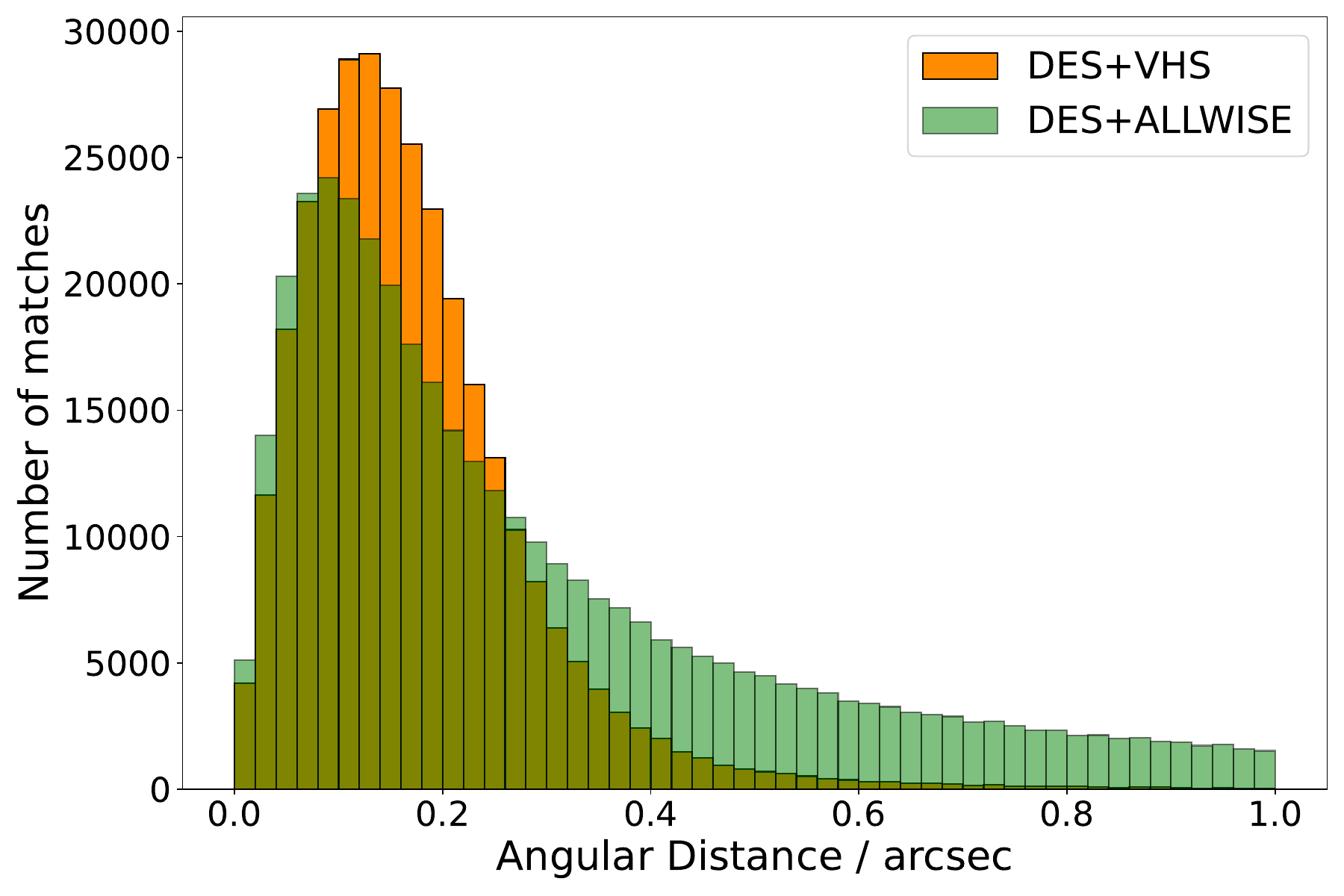}
    \caption{\centering The distribution of separations between DES and VHS sources in orange and DES and AllWISE sources in green.}
    \label{fig:angseparation}
\end{figure}

Figure \ref{fig:angseparation} shows the distribution of the angular separation of the matched sources between the DES and AllWISE (green) and the DES and VHS (orange) catalogues, providing some information on the average positional offsets and the matching accuracy. 

\begin{figure}[htbp]
    \centering
    \includegraphics[width=0.9\linewidth]{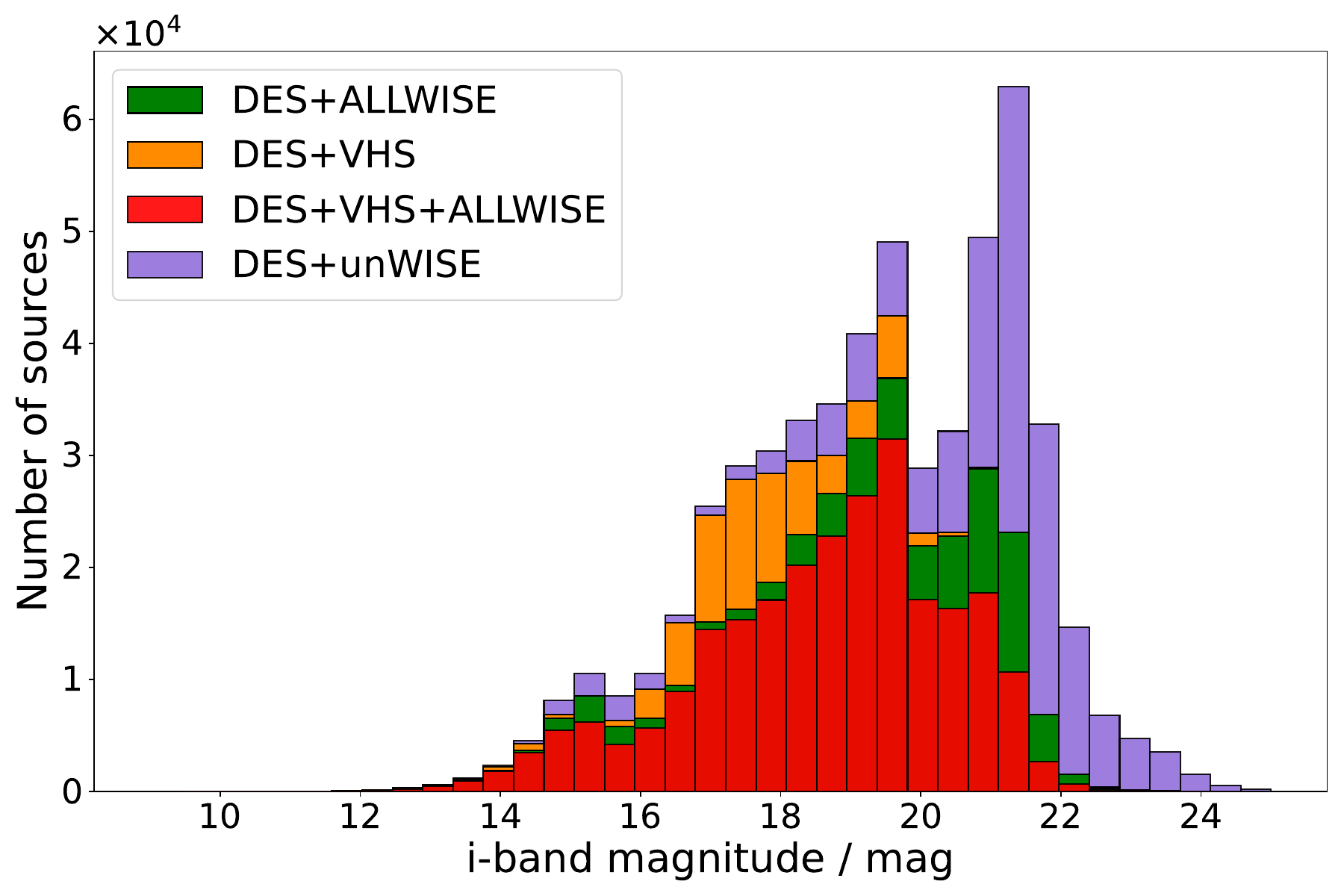}
    \caption{\centering Distribution of $i$-band magnitudes for the four matched catalogues, DES+AllWISE (green), DES+VHS (orange), DES+VHS+AllWISE (red) and DES+unWISE (purple).}
    \label{fig:imag}
\end{figure}

Figure \ref{fig:imag} presents the $i$-band magnitude distributions for the four catalogues, DES+AllWISE (green), DES+VHS (orange), DES+VHS+AllWISE (red) and DES+unWISE (purple). The distribution peak is around $i\approx 20$, and drops significantly beyond $i$ $\approx 21$ (except for DES+unWISE which uses a forced photometry association), suggesting that this is the effective limiting magnitude of the three matched catalogues. This highlights the relative depth and photometric characteristics of each catalogue in the $i$-band.

We focus on the $i$-band magnitude distribution since it provides an optimal balance between depth and sensitivity for galaxy detection. It is also less affected by atmospheric extinction and interstellar dust than bluer bands, and is accessible in the three catalogues, which allows for a comparison of their effective depths.

\subsubsection{Matching the complete Y6 Gold catalogue}
\label{sec:match_complete_gold}

For the comparison case of the information from the unWISE (deeper) dataset, we make use of the catalogue information from DECaLS DR10, and perform a similar positional matching\footnote{Using the tool at \url{https://github.com/rongpu/desi-truth-table}} between said catalogue and Y6 Gold.

\section{Photo-$z$ algorithm and quality metrics}

\subsection{The DNF algorithm}
For the analysis and determination of photometric redshifts, we used the \textbf{DNF \textit{(Galaxy photometric redshift by Directional Neighbourhood Fitting)}} algorithm developed by \cite{DNF}, which is the core estimator in DES Y6 Gold. This software, developed in Python, allows for the calculation of photometric redshifts by applying advanced fitting techniques, and provides three different distance metrics: Euclidean (ENF), Angular (ANF), and Directional (DNF). In particular, the ANF metric was employed for the estimation, using the code available at \href{https://github.com/ltoribiosc/DNF\_photoz}{\texttt{https://github.com/ltoribiosc/DNF\_photoz}}. 

In ANF, the photometric redshift of a galaxy is calculated based on the redshifts of nearby training galaxies within the multimagnitude space. Key factors that influence the accuracy of this estimation include how the neighbourhood is defined, the number of neighbours selected, and the method used to perform the local fit that relates magnitudes to redshift. The redshifts of the training sample define a hypersurface across the multimagnitude space. In this context, for any given point in this space, which corresponds to a galaxy with unknown redshift, the shape of the local hypersurface can vary in smoothness depending on the direction. Accurate redshift estimation is more likely when the neighbourhood is selected along a direction in which the redshift hypersurface varies smoothly, allowing for a better fit to the local structure.

To define the neighbourhood, ANF uses the normalised inner product definition (NIP) (\cite{Sanchez}), which is equivalent to the cosine of the angle, $\alpha$, between two multimagnitude vectors. The NIP metric identifies two galaxies as neighbours, and therefore as having a comparable redshift, when their relative distribution of magnitudes is alike, prioritizing similarity in colour over absolute proximity in magnitude space. The highest degree of similarity is achieved when the vectors are aligned ($\alpha = 0$), meaning the ratio between their photometric bands is preserved. In such cases, the objects are considered neighbours in the multimagnitude space.


The DNF algorithm provides the main photo-$z$ value as: \texttt{DNF$\_$Z} determined by the fit of a number (\texttt{DNF$\_$NNEIGHBORS}) of neighbouring galaxies to a hyperplane in magnitude space. 

\subsection{Metrics}
\label{sec:metrics}

The sample we are analysing includes magnitude measurements from the broad band filters of all three surveys, as well as the spectroscopic redshifts. Taking the spectroscopic measurement as the truth value for the redshift, we can evaluate some metrics for the quality of the photometric redshift estimation, described in the following subsections. 
 
\subsubsection{Bias}
    The bias, $\mu$: measures the systematic offset between $z_{phot}$ and $z_{spec}$, where $z_{phot}$ and  $z_{spec}$ refer to the photometric and spectroscopic redshift, respectively. Normally the individual residual for each galaxy is defined as $\Delta z'= z_{phot}-z_{spec}$. We decided however to employ a more robust estimate for bin $j$ through the median instead of the mean, and normalize it by (1 + $z_{spec}$) to provide a more homogeneous metric that gives the same fractional error in wavelength ($\Delta z/(1+z) = \Delta \lambda/\lambda$):
    \begin{equation}
        \mu = median(\Delta z_j),
    \end{equation}
    where $\Delta z_j$ is the collection of the $N$ values of $\Delta z$ in bin $j$, and $\Delta z = \Delta z' / (1 + z_{spec})$.

    The error is estimated as:
    \begin{equation}
        \Delta \mu = 1.253 \cdot \frac{NMAD}{\sqrt{N}},
    \end{equation}
   where NMAD is the normalized mean absolute deviation $NMAD = 1.48\cdot MAD$ \citep{kendall1946} under the assumption of a Gaussian distribution, and MAD is:
   \begin{equation}
       MAD = median(|\Delta z_j - \mu|).
   \end{equation}
   
\subsubsection{The $\sigma_{68}$ containment scatter estimate}

    Precision in a 68-quantile, $\sigma_{68}$ quantifies the scatter or dispersion of the photometric redshifts. It represents the width of the distribution of photometric redshifts around the median that contains 68$\%$ of the data points. Specifically, it corresponds to the 68$\%$ quantile error, and is defined as:  
    \begin{equation}
        \sigma_{68}= \frac{1}{2}(P_{84}-P_{16}),
    \end{equation} 
    where $P_{84}$ and $P_{16}$ are the 84th and 16th percentiles of the cumulative distribution of $\Delta z_j$, respectively, including the normalisation factor (1 + $z_{spec}$), as per the definition we are using. The value of $\sigma_{68}$ is indicative of how well the photometric redshift estimates are clustered around the spectroscopic redshift.  
        
    The normalisation is particularly useful for ensuring that the error does not systematically increase with redshift, otherwise the metric would exhibit an artificial redshift dependence.

    The error in this case is estimated as:

    \begin{equation}
        \Delta \sigma_{68} = \frac{\sigma_{68}}{\sqrt{2N}}.
    \end{equation}
    
\subsubsection{Outlier fraction} 

    This metric quantifies the fraction $f$ of objects with large biases, for which the photometric redshift estimates are significantly inaccurate. It is defined following the criterion of \cite{Banerji2015}, where an object is classified as an outlier if the following condition is satisfied:
    \begin{equation}
       |\Delta z|  > 0.15.
    \end{equation}
    This fixed-threshold definition is preferred over $\sigma_{68}$-based criteria (|$\Delta z| > n \cdot \sigma_{68}$), as the value of $\sigma_{68}$ tends to increase with redshift. Therefore, applying a threshold that scales with $\sigma_{68}$ can lead to an underestimation of the outlier rate in higher redshift bins. In contrast, the Banerji criterion sets a constant threshold for all redshift bins, allowing for stronger outlier identification and a direct comparison across the entire redshift range.

    The error is appropriately characterised from the error of a binomial distribution with probability $f$:

    \begin{equation}
        \Delta f = \sqrt{\frac{f(1-f)}{N}}.
    \end{equation}





\section{Results}
\label{sec:result}
In this section, we apply the metrics described in Section \ref{sec:metrics} to compare the photometric redshift estimates obtained from the DES dataset using different combinations.

First, we analyse the improvements in photometric redshift determination when applying DNF method in four catalogues: DES+WISE (AllWISE and unWISE versions), DES+VHS and DES+VHS+WISE. Next, we compare our results with the findings presented by \cite{Banerji2015} using early DES and VHS data.


\subsection{Comparative analysis}

We study how the results change within the same matched spectroscopic catalogue if we take into account for the DNF estimate only the optical bands from DES or combination with other infrared surveys. 

\begin{figure}[ht]
\centering
  \subfloat{
    \includegraphics[width=0.48\textwidth]{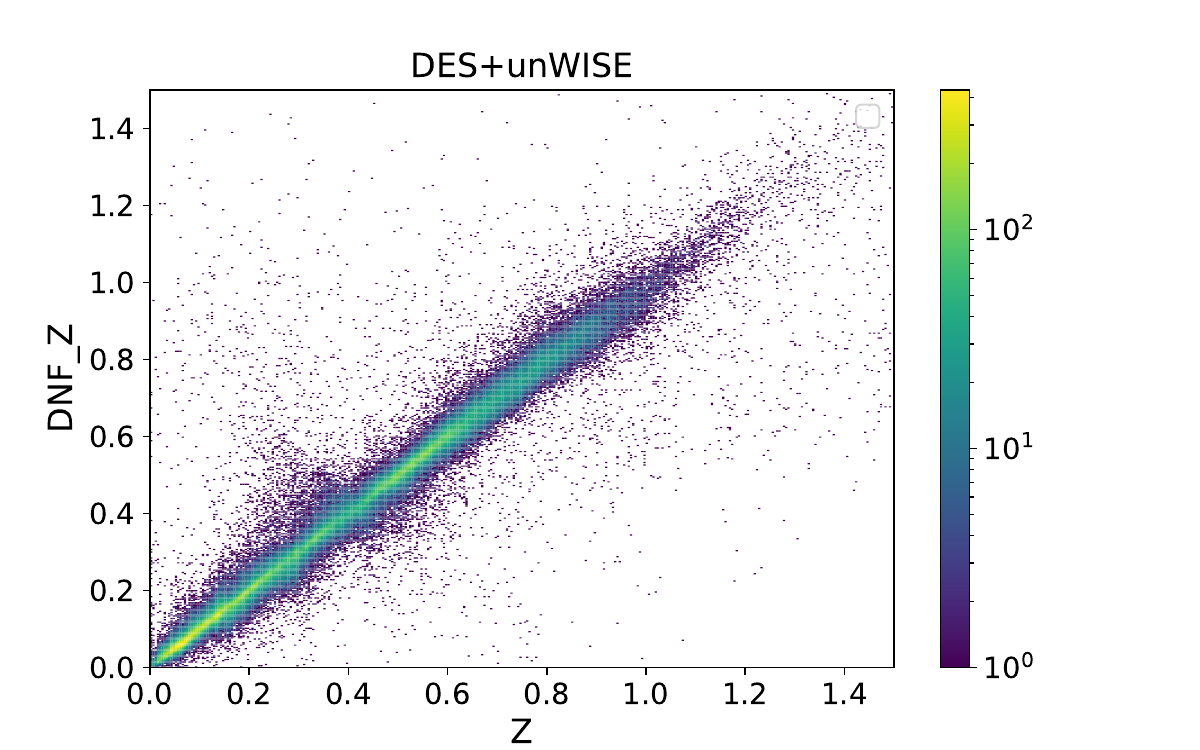}
  }
\caption{Photometric vs spectroscopic redshift for the DES+unWISE matched catalogue.}
\label{f:des+unwise_zspeczphot}  
\end{figure}

Figure \ref{f:des+unwise_zspeczphot} shows the comparison between the DNF estimate and spectroscopic redshifts for the DES+unWISE catalogue. 

\begin{figure}[ht]
  \centering  
  \subfloat[Median normalized bias, as a function of the spectroscopic redshift bin. For $z>1.4$ the blue (DES) and orange (DES+VHS) points are out of range, at $\Delta z< -0.15$.\label{f:all_bias_pz}]{
    \includegraphics[width=0.48\textwidth]{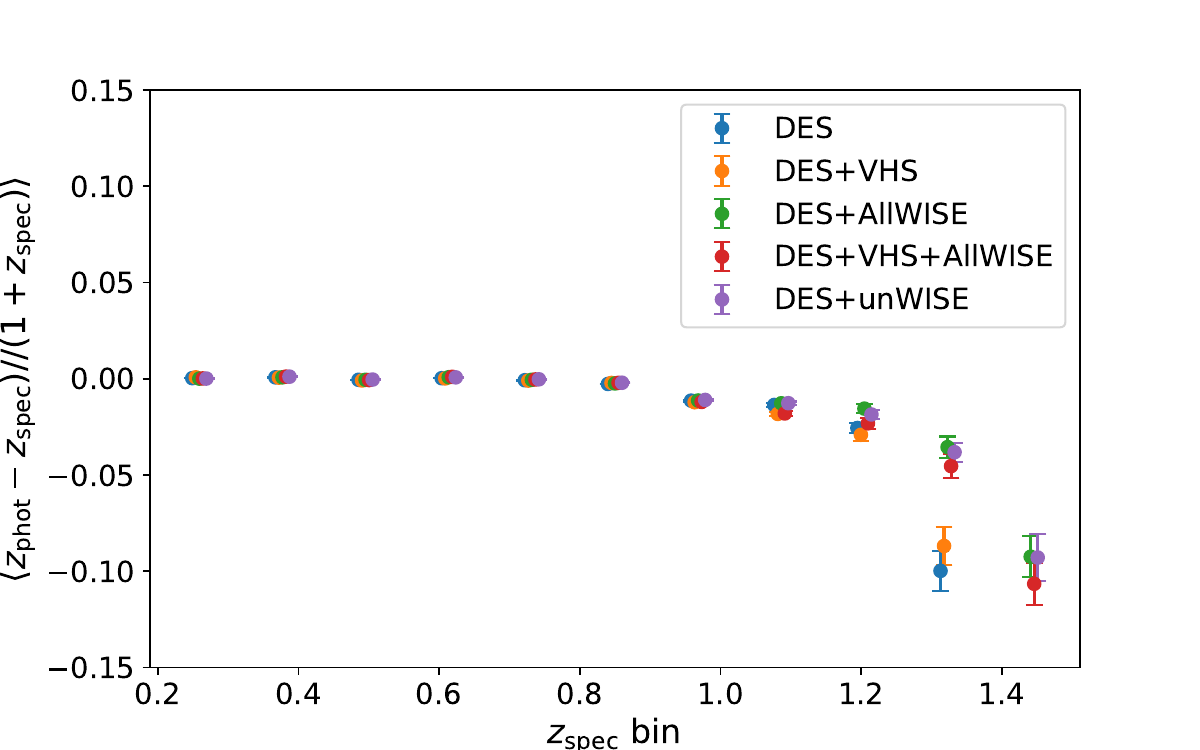}
  }
  
  \subfloat[The 68$\%$ quantile normalized error,  as a function of the spectroscopic redshift bin.\label{f:all_s68_pz}]{
    \includegraphics[width=0.48\textwidth]{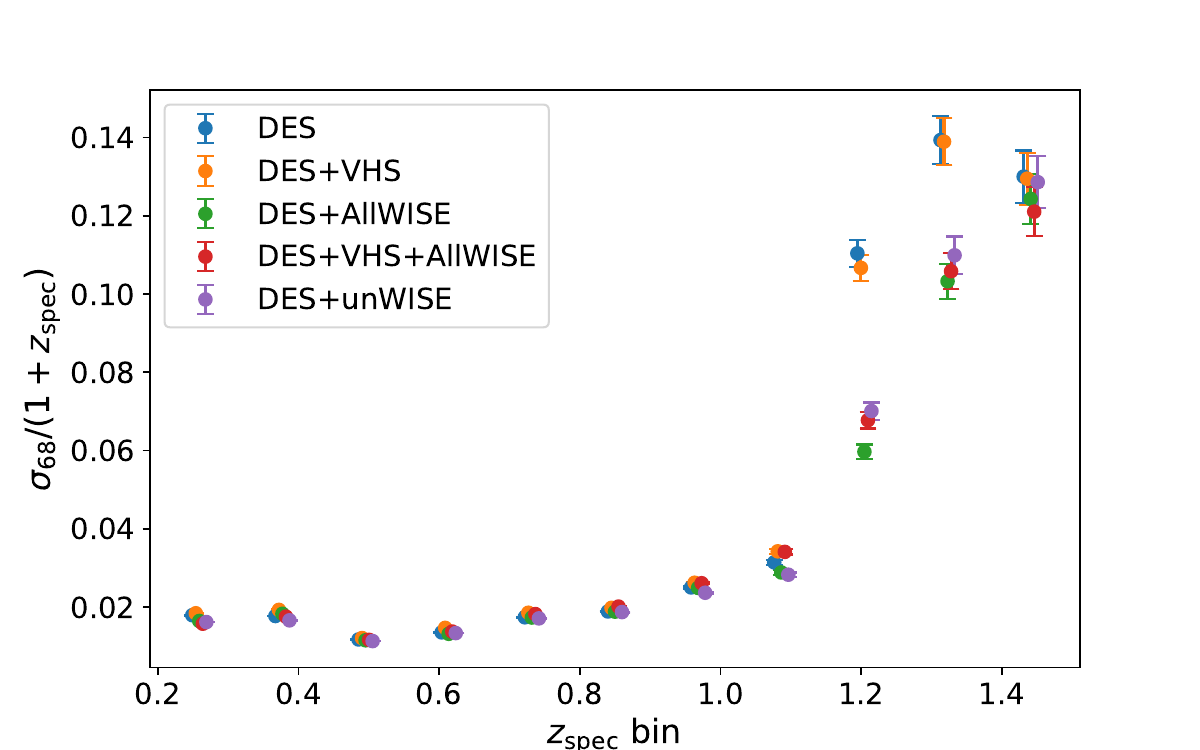}
  }
  
  \subfloat[Outlier fraction defined as objects with $\Delta$z > 0.15,  as a function of the spectroscopic redshift bin.\label{f:all_outliers_pz}]{
    \includegraphics[width=0.48\textwidth]{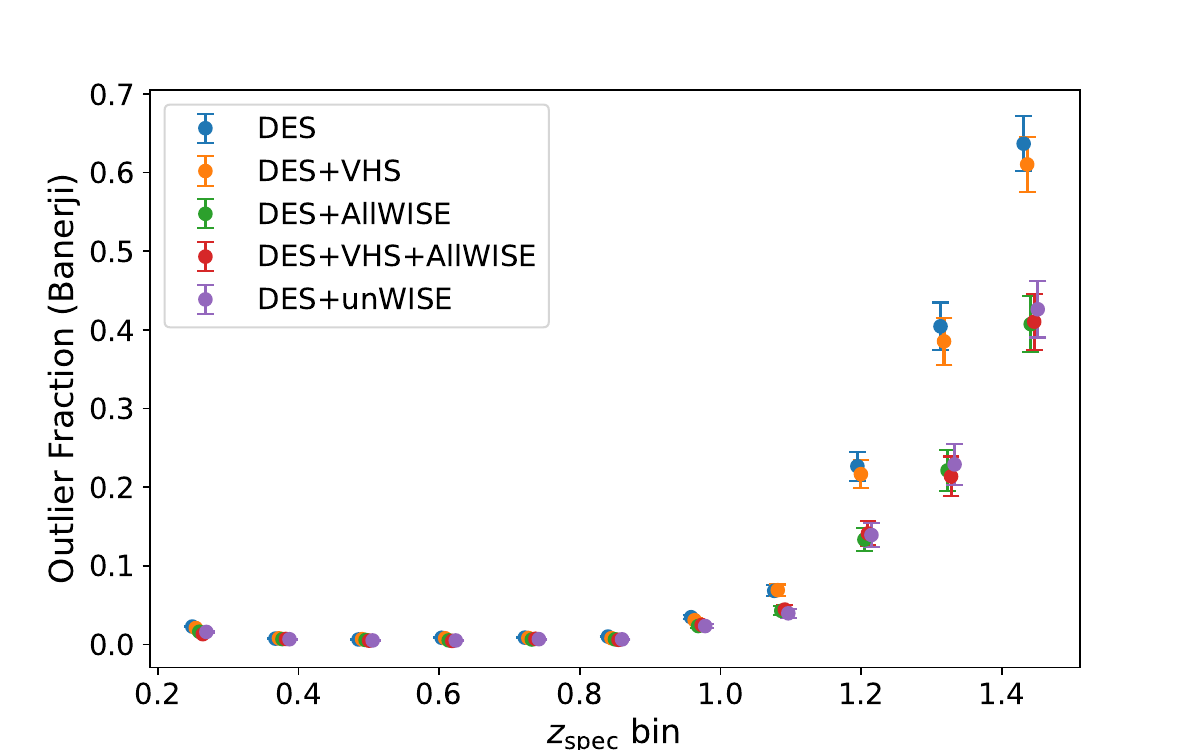}
  }

  \caption{Metrics for the matched catalogue as a function of photometric redshift bins.}
  \label{f:des+all_pz}
\end{figure}

\begin{figure}[ht]
  \centering
  \subfloat[Median normalized bias, as a function of the $i$ magnitude bin.\label{f:all_bias_imag}]{
    \includegraphics[width=0.48\textwidth]{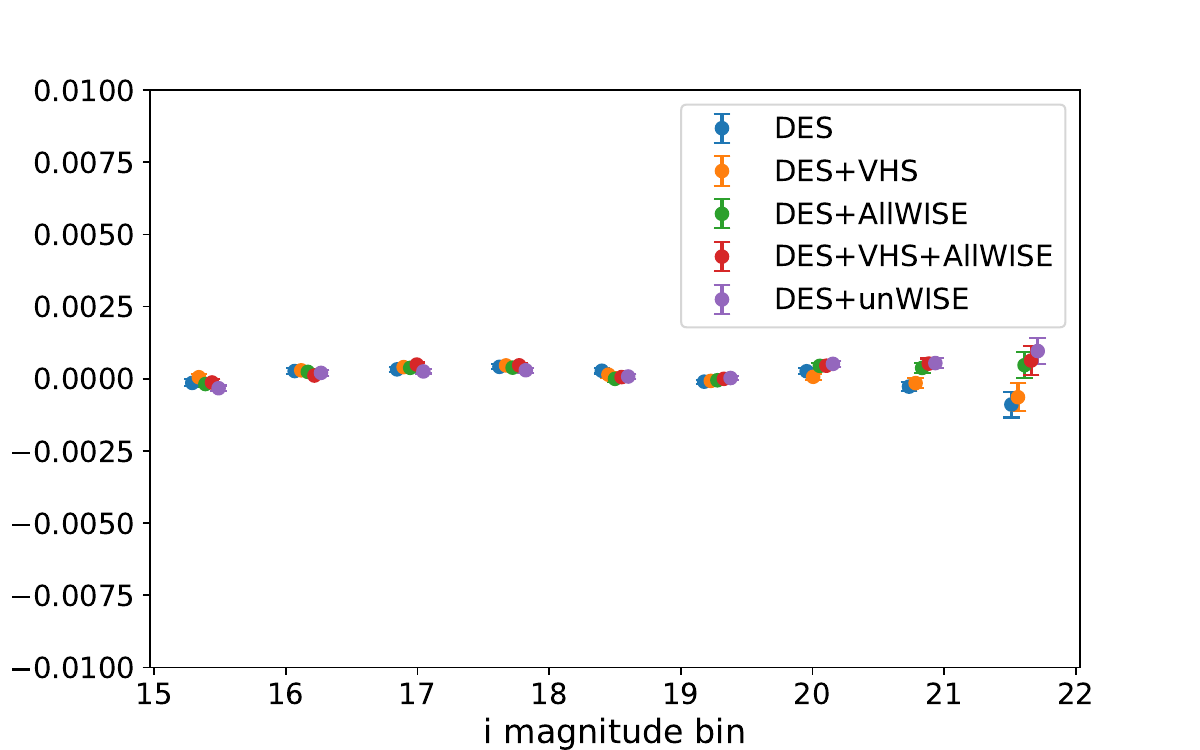}
  }
  
  \subfloat[The 68$\%$ quantile normalized error,  as a function of the $i$ magnitude bin.\label{f:all_s68_imag}]{
    \includegraphics[width=0.48\textwidth]{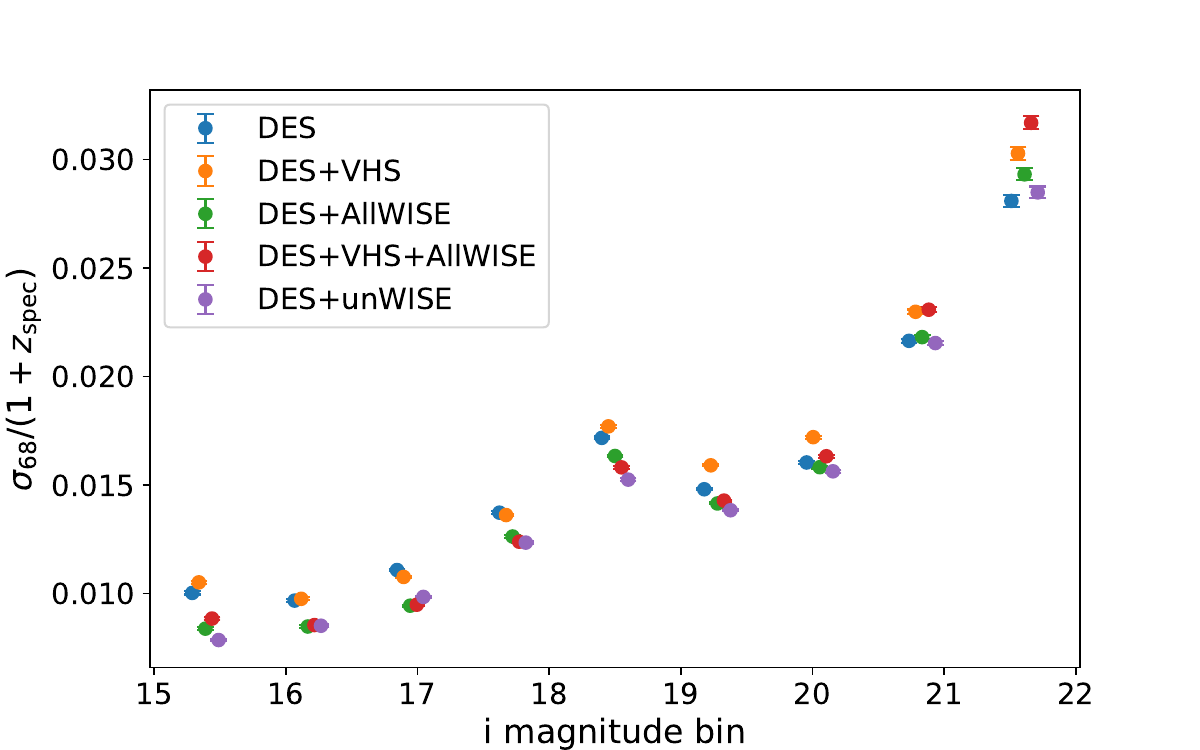}
  }
  
  \subfloat[Outlier fraction defined as objects with $\Delta$z > 0.15,  as a function of the $i$ magnitude bin.\label{f:all_outliers_imag}]{
    \includegraphics[width=0.48\textwidth]{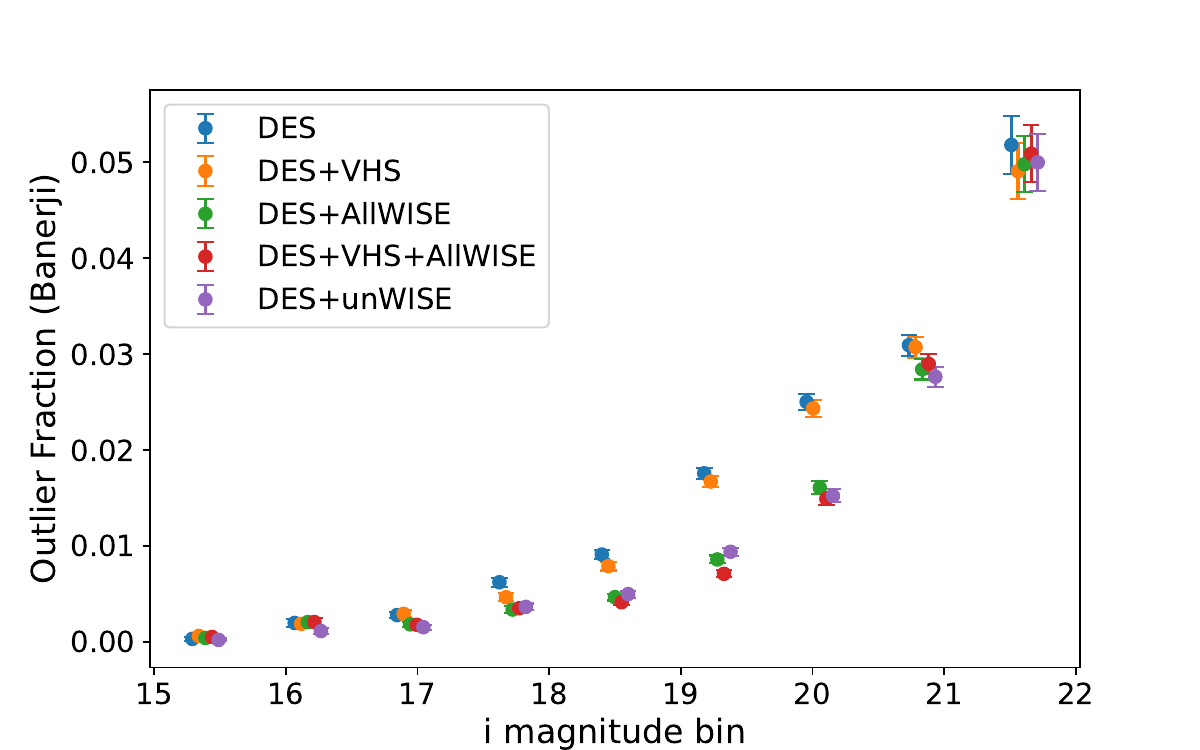}
  }

  \caption{Metrics for the matched catalogue as a function of i-band magnitude bins.}
  \label{f:des+all_mag}
\end{figure}

The metrics from Section \ref{sec:metrics} are presented for different combinations of the datasets enumerated in Section \ref{sec:datasets} in Figures \ref{f:des+all_pz} and \ref{f:des+all_mag}. 

The first result that is immediately noticeable is the improvement in the scatter and outlier rate, when infrared information is available. This is more evident at higher redshifts ($z > 1.1$), as one might expect, as the Balmer break moves into this wavelength regime. 

Moreover, the most important contribution to the improvement comes from the addition of mid-infrared (WISE) data. VHS bands ($J$ and $Ks$) only provide a marginal gain over the $grizY$ contribution in the studied redshift range. We did not include $H$ as this would restrict the amount of matches (and eventually, size of the enhanced catalogue) because of the lower sky coverage, but we checked that it did not provide any further enhancement. On the other hand, restricting our dataset to including $W3$ and $W4$ would limit the area of the catalogue severely. 

Interestingly, as a function of source brightness we find an overall improvement with infrared data, until the signal to noise for the latter is so low that there is no significant contribution over the deeper, optical information. All three metrics indicate that the photometric redshift performance deteriorates toward fainter magnitudes. In general, these faintest bins show larger median bias, scatter and outlier fractions. This behaviour reflects the reduced photometric signal-to-noise, which limits the constraints on spectral energy distributions and consequently degrades the photo-$z$ precision.

\begin{figure}[ht]
    \centering    \includegraphics[width=0.48\textwidth]{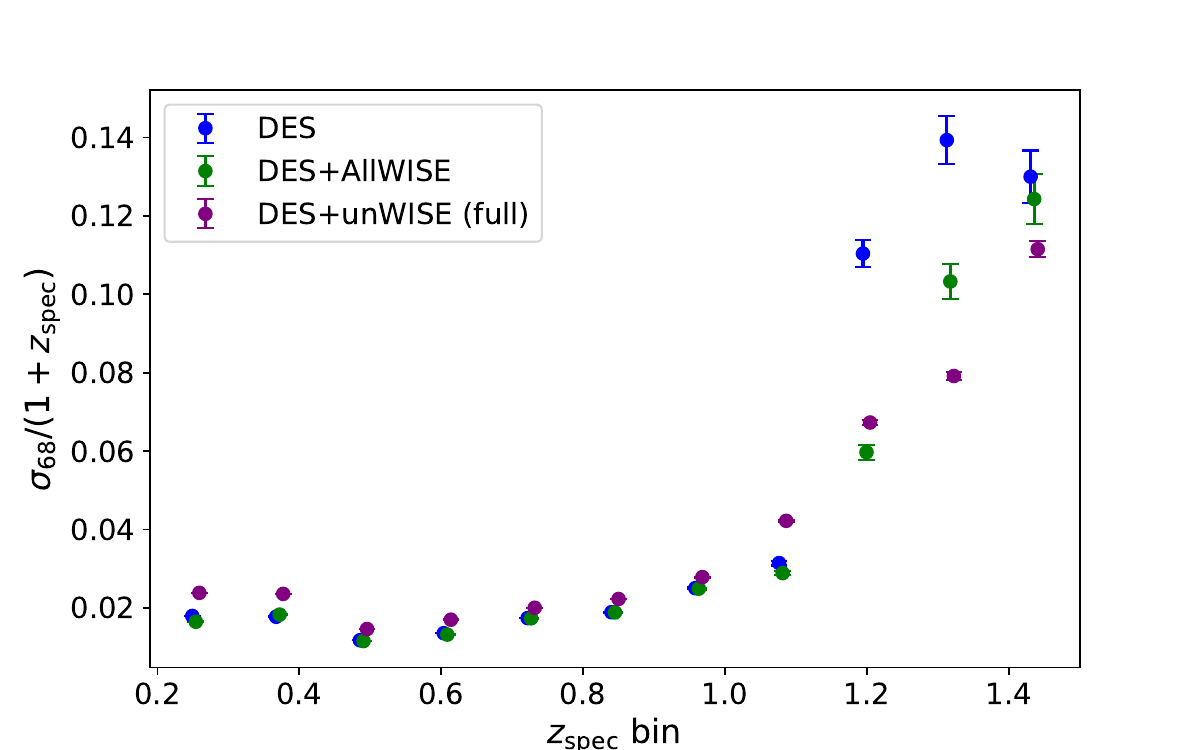}
    \caption{The $\sigma_{68}$ scatter shown comparatively for the optical and optical + AllWISE data, as in Figure \ref{f:all_s68_pz}, plus the metric for the matched spectroscopic catalogue using forced photometry from unWISE (deeper dataset).}
    \label{fig:des+unwisefull}  
\end{figure}

The plots don't show a significant difference either when comparing for the same set of galaxies, the AllWISE catalogue matches and the unWISE forced photometry associated fluxes (using DECaLS sources as proxies for DES objects). This gives us confidence that we can rely on the unWISE matches, at least until mid to bright magnitudes. In order to understand the behavior of the unWISE matches at larger depths, in  Figure \ref{fig:des+unwisefull} we overplot the $\sigma_{68}$ values comparing these two noting that they correspond to different catalogues. An overall degradation is observed at low to mid-redshifts, as expected due to the larger signal-to-noise from the catalogue corresponding to the blue and green points, whereas the purple catalogue contains many more matches thanks to forced photometry. At high redshift however this additional low signal to noise data does provide sufficient information to give an additional edge in the metrics.

We also experimented with removing the DES $Y$ band on account of having a poorer signal-to-noise on earlier DES analyses that pushed to avoiding this band altogether for photo-z estimates. It did not have a significant impact either on one direction or the other. 

Some of these quantities and trends can be compared numerically at Table \ref{tab:photoz_metrics}.

\begin{table*}
\caption{Photo-$z$ performance metrics for three redshift bins considered in this work.}
\label{tab:photoz_metrics}
\centering
\textbf{Bin $z = 0.2-1.0$}

\begin{tabular*}{\textwidth}{@{\extracolsep{\fill}}lccc}
\hline
Dataset & Bias & $\sigma_{68}$ & Outlier fraction  \\
\hline
DES & $-0.00027 \pm 0.00004$ & $0.01568 \pm 0.00003$ & $0.01169 \pm 0.00028$ \\
DES+VHS & $-0.00032 \pm 0.00005$ & $0.01653 \pm 0.00003$ & $0.01103 \pm 0.00027$ \\
DES+AllWISE & $-0.00017 \pm 0.00004$ & $0.01533 \pm 0.00003$ & $0.00862 \pm 0.00024$ \\
DES+VHS+AllWISE & $-0.00007 \pm 0.00005$ & $0.01547 \pm 0.00003$ & $0.00774 \pm 0.00023$ \\
DES+unWISE & $-0.00005 \pm 0.00004$ & $0.01502 \pm 0.00003$ & $0.00843 \pm 0.00024$ \\
\hline
\end{tabular*}

\textbf{Bin $z = 1.0-1.2$}

\begin{tabular*}{\textwidth}{@{\extracolsep{\fill}}lccc}
\hline
Dataset & Bias & $\sigma_{68}$ & Outlier fraction  \\
\hline
DES & $-0.01494 \pm 0.00073$ & $0.03454 \pm 0.00052$ & $0.08425 \pm 0.00596$ \\
DES+VHS & $-0.01917 \pm 0.00081$ & $0.03759 \pm 0.00057$ & $0.08157 \pm 0.00588$ \\
DES+AllWISE & $-0.01401 \pm 0.00073$ & $0.03124 \pm 0.00047$ & $0.05292 \pm 0.00480$ \\
DES+VHS+AllWISE & $-0.01900 \pm 0.00085$ & $0.03509 \pm 0.00053$ & $0.05430 \pm 0.00486$ \\
DES+unWISE & $-0.01372 \pm 0.00069$ & $0.02945 \pm 0.00045$ & $0.05062 \pm 0.00470$ \\
\hline
\end{tabular*}

\textbf{Bin $z = 1.2-1.5$}

\begin{tabular*}{\textwidth}{@{\extracolsep{\fill}}lccc}
\hline
Dataset & Bias & $\sigma_{68}$ & Outlier fraction  \\
\hline
DES & $-0.11388 \pm 0.00751$ & $0.14297 \pm 0.00401$ & $0.43956 \pm 0.01967$ \\
DES+VHS & $-0.10680 \pm 0.00708$ & $0.14127 \pm 0.00396$ & $0.41601 \pm 0.01953$ \\
DES+AllWISE & $-0.04654 \pm 0.00400$ & $0.12104 \pm 0.00339$ & $0.25472 \pm 0.01728$ \\
DES+VHS+AllWISE & $-0.05609 \pm 0.00436$ & $0.11821 \pm 0.00331$ & $0.25589 \pm 0.01729$ \\
DES+unWISE & $-0.04706 \pm 0.00388$ & $0.12136 \pm 0.00340$ & $0.26845 \pm 0.01756$ \\
\hline
\end{tabular*}
\end{table*}

The lack of the $u$-band in DES measurements limits precision at low redshifts, where ultraviolet information is crucial to break colour-redshift degeneracies, so the increase of the scatter parameter is explained. Furthermore, at high redshifts, the error bars also increases, due to the decreasing number of galaxies with spectroscopic redshifts in this regime. Moreover, the high-z sample is biased towards brighter galaxies, which are more easily detected.  

The outlier fraction remains below 5$\%$ for $z<0.9$, and increases when we consider high redshifts,  due to the combined effects of reduced training density and intrinsic photometric uncertainties. It shows a high rise for redshift bins larger than 1.0 in the optical-only data, reflecting the degradation of photo-$z$ reliability in the absence of infrared constraints. Nevertheless, we observe a significant reduction in the outlier rate when combining DES and WISE data, which confirms that the inclusion of near-infrared bands constrains redshift estimates, especially in the high-z regime.

\subsection{Comparison with other works}

\subsubsection{DES SV + VHS from Banerji et al.}

To contextualize the results obtained in this work, we compare them with those from \cite{Banerji2015}, who conducted a preliminary analysis using DES science verification data and VHS photometry with very early versions of the datasets and smaller statistics. 

The addition of VHS near-infrared data to DES photometry resulted in a modest improvement in performance: the scatter decreased from 0.058 to 0.052, reaching its lowest values around redshift 0.7, where the 4000 \AA \ break lies within the optical bands. The outlier fraction also decreased from 13 to 10$\%$. Their analysis emphasized that the inclusion of NIR bands improves the robustness of the photometric redshift metrics, especially the scatter parameter, $\sigma_{68}$. VHS information helped reduce degeneracies, particularly between early-type galaxies and reddened starbursts, by better constraining galaxy spectral energy distributions (SEDs). 

\begin{figure}[ht]
    \centering   
    \subfloat[Bias as a function of photometric redshift. \label{fig:banerji_comparison_bias}]{\includegraphics[width=0.48\textwidth]{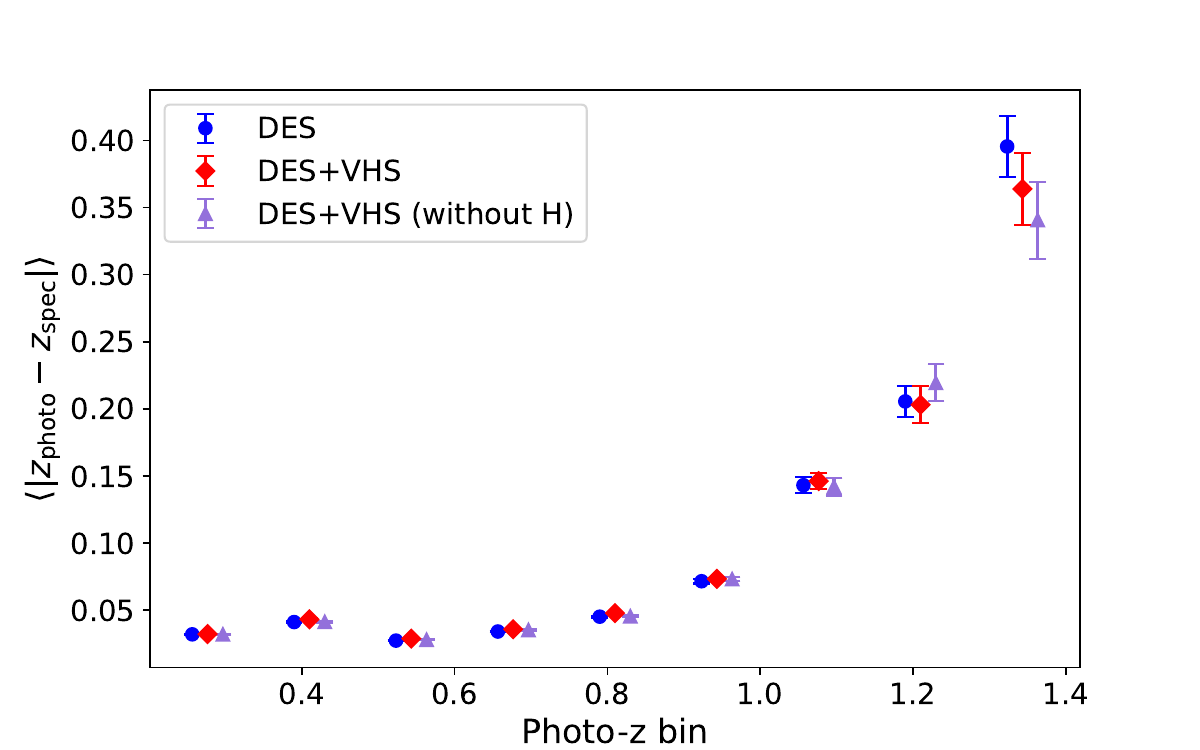}}
    
    \subfloat[Scatter as a function of photometric redshift. \label{fig:banerji_comparison_scatter}]{\includegraphics[width=0.48\textwidth]{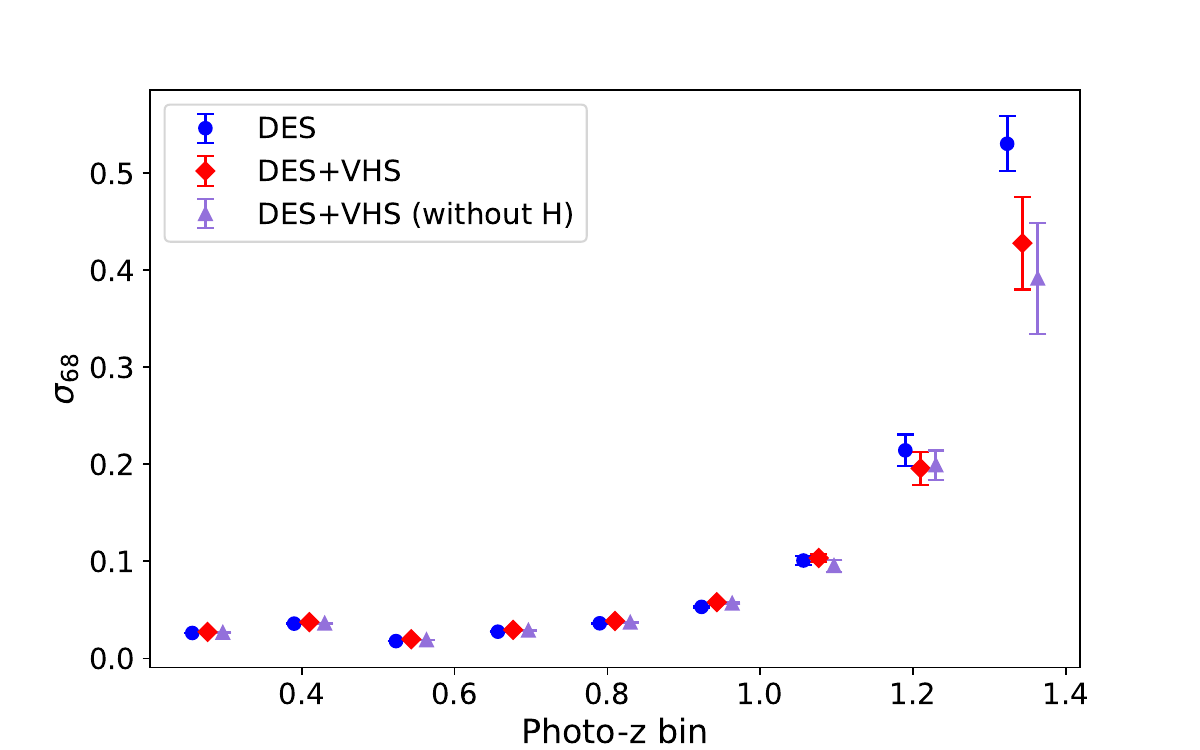}}
    
    \caption{Metrics using combination with VHS data, as shown in \cite{Banerji2015}}
    \label{fig:banerji_comparison}  
\end{figure}

When comparing the results obtained by \cite{Banerji2015} in their Figure 10 and those presented in this work (Figure \ref{fig:banerji_comparison}), we observe a moderate agreement with their conclusions. As a function of \textit{photometric} redshift, there is a slight improvement above $z > 1.2$. We opted, however, to use spectroscopic redshifts as a reference as it would correspond to a measure of the real metric for farther away sources. While \cite{Banerji2015} employed a template-fitting technique based on synthetic galaxy spectra, \textsc{lephare} (\cite{lephare}), this work is based on a training-based approach, specifically the DNF algorithm (\cite{DNF}). In addition, the input datasets vary. \cite{Banerji2015} utilised a combination of DES science verification data and VHS photometry, and in this work, we use a newly merged catalogue including updated DES data and extended infrared coverage from VHS. These differences in the photometric input, including depth, photometric uncertainties, sample completeness, and intrinsic characteristics of the machine learning model can lead to systematic differences in the performance metrics. The size of the training sample is a factor that influences the quality of the photo-$z$s. The outlier fraction in our work remains below 5$\%$ for z$<$0.9, which is substantially better than the 10–13$\%$ range reported by \cite{Banerji2015}. But the overall conclusion remains the same: the $JHKs$ bands provide modest improvements at the available VHS and DES depths.

Concerning other aspects of the analyses, both studies are limited by the lack of photometric coverage at low redshifts, mainly due to the absence of the $u$-band in the DES dataset. This band is essential for constraining spectral features such as the Balmer break, which plays a key role in improving redshift estimates for nearby galaxies. While \cite{Banerji2015} explicitly mentions this issue in the context of their template-based method, our training-based approach is similarly affected. Although the inclusion of infrared data (e.g., from VHS or WISE) helps to improve the estimates, especially at intermediate redshifts, it does not fully compensate for the missing information at low redshift.

Regarding the results, the use of a training-based method leads to a clear improvement in the photometric redshift scatter at low redshifts, as indicated by systematically lower values of compared to the template-based estimates. However, at higher redshifts (above z=1.2), a slight degradation in performance is observed, manifested by an increase in the last redshift bin. This is probably caused by the absence of high-redshift galaxies in the training sample, which typically degrades the model's predictive accuracy in this regime. In addition, differences in the depth and photometric quality of the datasets used, as well as in the different methods used for photometric error treatment and feature selection, can potentially be responsible for the differences noted at high redshift. Nevertheless, the general trends are consistent with previous observations: the addition of infrared photometry improves photometric redshift estimations, particularly by reducing the scatter and bias at high and intermediate redshifts. These results support the utility of combining infrared and optical data and also show the potential of using machine learning techniques to enhance photometric redshift estimation in future large-scale surveys.

\subsubsection{Rubin + Euclid from Zhang et al.}

Recently, \cite{dp1pzs} presented for the first time the combination of Vera Rubin Observatory's preliminary data (Data Preview 1) with a similar, early dataset from the Euclid space observatory, using $ugrizy$ and $YJH$ filters respectively, up to a redshift of $z \sim 2$. Their conclusions are broadly consistent with those presented here in the redshift range $z < 1.5$, where the inclusion of infrared data up to $2 \mu m$ is not apparent. However it starts becoming much more noticeable beyond this range, where the DES data is not deep enough to provide any insight on this regard.

\section{Conclusions}

In this work, we present a new value-added version of the DES Y6 Gold catalogue including Directional Neighbourhood Fitting (DNF) photometric redshifts using forced photometry from unWISE and corresponding DECaLS DR10 catalogue matches, in particular, using the $W1$ and $W2$ bands. 

This choice for the infrared-enhanced version of the catalogue is the result of the analysis done on the spectroscopic matching for a subset of galaxies from DES. Our results indicate that the addition of infrared data from both VHS and WISE surveys, can help reduce the photometric redshift scatter, though only marginally in the case of VHS. Adding infrared data reduces systematic bias and improves the overall photometric redshift performance, particularly at higher redshifts in comparison with only using optical data from DES (by more than $50\%$ in bias, $15\%$ in scatter and reducing outliers to a more manageable $25\%$).  These results confirm that, both through template-based methods (such as those used by \cite{Banerji2015}) and through machine learning techniques, with a different code (DNF), the inclusion of IR data leads to improvements in photometric redshift estimation, furthermore, the training-based approach used here achieves even lower values of bias and scatter, especially at low to intermediate redshifts. Moreover, the obtained results are globally better for the DES+WISE sample, with improvements that are statistically significant. The combined use of DES, VHS, and WISE data does not significantly improve the results beyond what is achieved using DES+WISE alone, suggesting that WISE data may already capture the key IR features needed for accurate photo-$z$ estimation with DNF, at the DES and VHS depths explored here. Given that VHS also limits further the coverage area, we decide to publish a new catalogue based on matches to WISE data (the unWISE version).

It is worth noting that the inclusion of infrared data opens up the opportunity of improving star-galaxy separation as demonstrated for the DES Y6 \texttt{MAGLIM++} sample in \cite{2026arXiv260504220W} and shown comparatively in \cite{2018MNRAS.481.5451S}.

With these results in hand, we have created an updated version of the DES Y6 Gold catalogue, available at the CosmoHub platform\footnote{\url{https://cosmohub.pic.es}}, that we describe in Appendix \ref{app:y6goldir}. The metrics obtained in Section \ref{sec:result} are not directly extrapolable to this data set strictly, due to differences in colour coverage between the training set and the overall sample (as with the metrics shown in \cite{y6gold}), but they indicate that the inclusion of infrared data improves the overall photo-$z$ point estimate performance. 

Finally, these efforts are highly relevant in the context of upcoming wide-area imaging large-scale surveys such as the Vera C. Rubin Observatory’s LSST and Euclid, which will demand precise and accurate photometric redshifts to unlock the next generation of cosmological and astrophysical discoveries. 

\begin{acknowledgements}
ISN would like to acknowledge N.Weaverdyck for pointers to the usage of DECaLS data. MMP acknowledges grant PID2024-155875OB-I00 funded by MICIU/AEI/10.13039/501100011033/FEDER, EU. ISN, JdV and LT acknowledge grants PID2021-123012NB-C42 and PID2024-159420NB-C42 from MICINN/MCIU/AEI in Spain. JC acknowledges support from the grant PID2024-159420NB-C44, funded by by MICIU/AEI/10.13039/501100011033 and by ERDF/EU.
      
This work has made use of CosmoHub, developed by PIC (maintained by IFAE and CIEMAT) in collaboration with ICE-CSIC. It received funding from the Spanish government (grant EQC2021-007479-P funded by MCIN/AEI/10.13039/501100011033), the EU NextGeneration/PRTR (PRTR-C17.I1), and the Generalitat de Catalunya.

This project used data obtained with the Dark Energy Camera (DECam), which was constructed by the Dark Energy Survey (DES) collaboration. 

Funding for the DES Projects has been provided by the U.S. Department of Energy, the U.S. National Science Foundation, the Ministry of Science and Education of Spain, the Science and Technology Facilities Council of the United Kingdom, the Higher Education Funding Council for England, the National Center for Supercomputing Applications at the University of Illinois at Urbana-Champaign, the Kavli Institute of Cosmological Physics at the University of Chicago, Center for Cosmology and Astro-Particle Physics at the Ohio State University, the Mitchell Institute for Fundamental Physics and Astronomy at Texas A\&M University, Financiadora de Estudos e Projetos, Fundaç\~ao Carlos Chagas Filho de Amparo, Financiadora de Estudos e Projetos, Fundaç\~ao Carlos Chagas Filho de Amparo a Pesquisa do Estado do Rio de Janeiro, Conselho Nacional de Desenvolvimento Cientifico e Tecnologico and the Ministerio da Ciencia, Tecnologia e Inovaç\~ao, the Deutsche Forschungsgemeinschaft and the Collaborating Institutions in the Dark Energy Survey. The Collaborating Institutions are Argonne National Laboratory, the University of California at Santa Cruz, the University of Cambridge, Centro de Investigaciones Energeticas, Medioambientales y Tecnologicas-Madrid, the University of Chicago, University College London, the DES-Brazil Consortium, the University of Edinburgh, the Eidgenossische Technische Hochschule (ETH) Zurich, Fermi National Accelerator Laboratory, the University of Illinois at Urbana-Champaign, the Institut de Ciencies de l’Espai (IEEC/CSIC), the Institut de Fisica d’Altes Energies, Lawrence Berkeley National Laboratory, the Ludwig Maximilians Universitat Munchen and the associated Excellence Cluster Universe, the University of Michigan, NSF’s NOIRLab, the University of Nottingham, the Ohio State University, the University of Pennsylvania, the University of Portsmouth, SLAC National Accelerator Laboratory, Stanford University, the University of Sussex, and Texas A\&M University.

This publication makes use of data products from the Wide-field Infrared Survey Explorer, which is a joint project of the University of California, Los Angeles, and the Jet Propulsion Laboratory/California Institute of Technology, funded by the National Aeronautics and Space Administration, VHS, based on data products from observations made with ESO Telescopes at the La Silla or Paranal Observatories under ESO programme ID 179.A-2010 and DES, based in part on observations at Cerro Tololo Inter-American Observatory, National Optical Astronomy Observatory, which is operated by the Association of Universities for Research in Astronomy (AURA) under a cooperative agreement with the National Science Foundation.

The Legacy Surveys consist of three individual and complementary projects: the Dark Energy Camera Legacy Survey (DECaLS; Proposal ID 2014B-0404; PIs: David Schlegel and Arjun Dey), the Beijing-Arizona Sky Survey (BASS; NOAO Prop. ID 2015A-0801; PIs: Zhou Xu and Xiaohui Fan), and the Mayall z-band Legacy Survey (MzLS; Prop. ID 2016A-0453; PI: Arjun Dey). DECaLS, BASS and MzLS together include data obtained, respectively, at the Blanco telescope, Cerro Tololo Inter-American Observatory, NSF’s NOIRLab; the Bok telescope, Steward Observatory, University of Arizona; and the Mayall telescope, Kitt Peak National Observatory, NOIRLab. Pipeline processing and analyses of the data were supported by NOIRLab and the Lawrence Berkeley National Laboratory (LBNL). The Legacy Surveys project is honored to be permitted to conduct astronomical research on Iolkam Du’ag (Kitt Peak), a mountain with particular significance to the Tohono O’odham Nation.

NOIRLab is operated by the Association of Universities for Research in Astronomy (AURA) under a cooperative agreement with the National Science Foundation. LBNL is managed by the Regents of the University of California under contract to the U.S. Department of Energy.

BASS is a key project of the Telescope Access Program (TAP), which has been funded by the National Astronomical Observatories of China, the Chinese Academy of Sciences (the Strategic Priority Research Program “The Emergence of Cosmological Structures” Grant XDB09000000), and the Special Fund for Astronomy from the Ministry of Finance. The BASS is also supported by the External Cooperation Program of Chinese Academy of Sciences (Grant 114A11KYSB20160057), and Chinese National Natural Science Foundation (Grant 12120101003, 11433005).

The Legacy Survey team makes use of data products from the Near-Earth Object Wide-field Infrared Survey Explorer (NEOWISE), which is a project of the Jet Propulsion Laboratory/California Institute of Technology. NEOWISE is funded by the National Aeronautics and Space Administration.

The Legacy Surveys imaging of the DESI footprint is supported by the Director, Office of Science, Office of High Energy Physics of the U.S. Department of Energy under Contract No. DE-AC02-05CH1123, by the National Energy Research Scientific Computing Center, a DOE Office of Science User Facility under the same contract; and by the U.S. National Science Foundation, Division of Astronomical Sciences under Contract No. AST-0950945 to NOAO.

\end{acknowledgements}

\bibliography{biblio}
\begin{appendix}
\section{DES Y6 Gold IR catalogue description.}
\label{app:y6goldir}

The Y6 Gold IR is a subset of the official DES Y6 Gold catalogue \citep{y6gold} made available as a final legacy product by DES. This new catalogue is being made available upon publication in the CosmoHub platform\footnote{\url{https://cosmohub.pic.es/catalogs/414}} and includes a subset of the rows (633M objects instead of 691M) covering the same area, as a consequence of using the matched WISE forced photometry information obtained from the DECaLS DR10 release objects, which have been used as a proxy for DES catalogue objects. The  new catalogue contains the columns described in Table \ref{tab:catalog_columns}, most of them being identical to the parent objects in the Y6 Gold catalogue and DECaLS DR10 catalogue, as indicated in the corresponding entry. The new columns are DNF outputs with suffix 'IR' though the association of $W1W2$ data is also an collateral benefit for users of the DES dataset. Note that, in addition to any of the regular Y6 Gold selections recommended in \cite{y6gold}, it would be advisable to consider an additional cut in \texttt{WISEMASK}, which controls possible artifacts from the unWISE dataset. In addition, users are encouraged to consider more accurate layers of masking from the LRG veto mask, as described in \cite{2023AJ....165...58Z} and available at \url{https://data.desi.lbl.gov/public/ets/vac/lrg_veto_mask/v1/}

\begin{figure*}[ht]
    \centering    \includegraphics[width=0.90\textwidth]{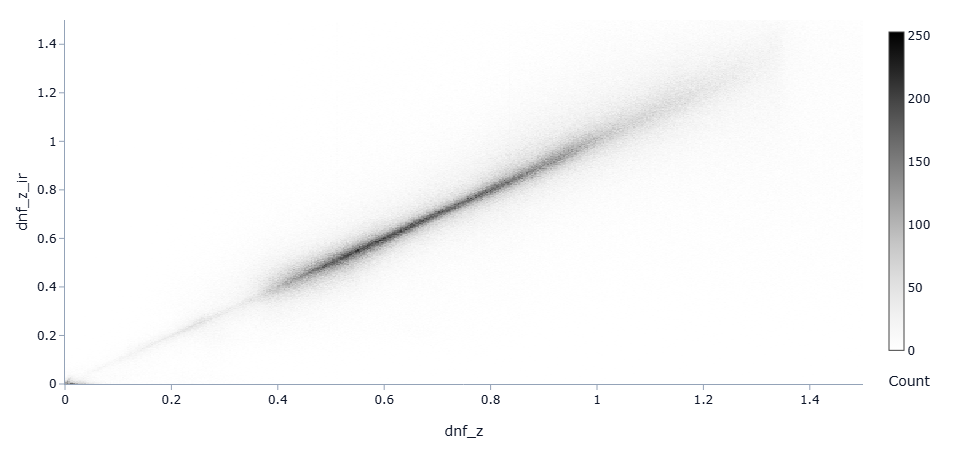}
    \caption{Sample exploratory plot showing the relation between the original $grizY$ DNF photo-z estimate from Y6 Gold and the new one generated for this work. Shown for illustration purposes of the tooling available at CosmoHub,}
    \label{fig:zvsz}  
\end{figure*}

The catalogue can be accessed through queries using the CosmoHub platform and linked to additional object information from the official DES Y6 Gold catalogue through the \texttt{COADD\_OBJECT\_ID} identifier, as detailed in the website documentation. Exploratory plots (such as the one shown in Figure\ref{fig:zvsz} can be obtained with the CosmoHub interface in seconds).

\begin{table*}
\caption{Columns from the DES Y6 Gold IR (infrared) catalogue.}
\label{tab:catalog_columns}
\centering
\small
\begin{tabularx}{\textwidth}{l l X}
\toprule
Column family & Units & Description \\
\midrule

COADD\_OBJECT\_ID 
& 
& Unique identifier for a Y6 coadd object. \\

RA, DEC 
& Degrees 
& Equatorial coordinates. \\

BDF\_MAG\_(G/R/I/Z/Y)\_CORRECTED
& Magnitudes $^{-1}$
& Fluxes corrected for Galactic extinction and PSF aperture ratio effects. \\

BDF\_MAG\_ERR\_(G/R/I/Z/Y)
& Magnitudes $^{-1}$
& Estimated uncertainties for BDF/GAP photometric quantities. \\

FLUX\_(W1/W2)
& nanomaggie
& WISE model flux in W1 (AB system) \\

FLUX\_IVAR\_(W1/W2)
& 1/nanomaggie$^2$
& Inverse variance of flux\_W1/W2 (AB system) \\

MW\_TRANSMISSION\_(W1/W2)
& 
& Galactic transmission in W1
 filter in linear units [0, 1] \\

FRACFLUX\_(W1/W2)
&
& Profile-weighted fraction of the flux from other sources divided by the total flux in W1 \\

WISEMASK\_(W1/W2)
& 
& W1/W2 bitmask as described in \url{https://www.legacysurvey.org/dr10/bitmasks} \\

MAG\_(W1/W2)
& magnitude
& Magnitudes computed from FLUX\_W1/W2 \\

MAG\_(W1/W2)\_CORR
& magnitude
& MAG\_W1/W2 corrected by MW\_TRANSMISSION\_W1/W2  \\

MAGERR\_(W1/W2)
& magnitudes
& W1/W2 magnitude errors (confirm) \\

DNF\_(Z/ZN)
&
& DNF photometric redshift estimate using either neighborhood fitting or nearest-neighbor estimation (using GRIZY) \\

DNF\_(Z/ZN)\_IR
&
& DNF photometric redshift estimate using either neighborhood fitting or nearest-neighbor estimation (using GRIZY) \\

DNF\_ZSIGMA\_IR
&
& Total DNF photo-$z$ uncertainty estimate. \\

DNF\_NNEIGHBORS\_IR
&
& Number of neighbors used in the DNF fit. \\

DNF\_ZERR\_PARAM\_IR
&
& Contribution to DNF uncertainty from photometric errors. \\

DNF\_ZERR\_FIT\_IR
&
& Contribution to DNF uncertainty from fit residuals. \\

\bottomrule
\end{tabularx}

\vspace{1mm}

\begin{minipage}{\textwidth}
\footnotesize
Names in parentheses indicate alternative column options separated by slashes.
\end{minipage}

\end{table*}

\end{appendix}
\end{document}